\documentclass[a4paper,10pt]{article}
\usepackage{graphicx}
\usepackage{subfigure}

\newcommand{\Ohm}{$\Omega$}

\begin{document}

\title{The effect of changing electrode metal on solution-processed flexible titanium dioxide memristors}

\author{Ella Gale$^{1,2}$, David Pearson$^{3}$, Steve Kitson$^{3}$ ,\\ Andrew Adamatzky$^{2}$ and Ben de Lacy Costello$^{1}$ \\
1 Department of Applied Sciences, University of the West of England, \\Frenchay Campus, Bristol, BS16 1QY\\
2 Bristol Robotics Laboratory, Bristol, BS16 1QY \\
3 Hewlett Packard Long Down Ave, Stoke Gifford, Bristol BS34 8QZ\\}

\maketitle

\begin{abstract}
Flexible solution processed memristors show different behaviour dependent on the choice of electrode material. Use of gold for both electrodes leads to switchable WORM (Write Once Read Many times) resistive devices. Use of aluminium for both electrodes increases the richness of behaviour allowing both curved and triangular memristive switching resistance memories. A comparison device with an aluminium bottom electrode and gold top electrode only exhibited significant memristive resistance switching when the aluminium electrode was the cathode, suggesting that the electrode is acting as a source/sink of oxygen anions. When the gold electrode was the cathode this electrode was deformed by oxygen evolution. These results demonstrate that aluminium is helpful for stabilising and promoting memristive behaviour in sol-gel TiO$_2$ devices and that changing electrodes from aluminium to gold creates fundamentally different device characteristics.
\end{abstract}

\vspace{2pc}

\bibliographystyle{unsrt} 

\section{Introduction}                       

Memristors are the fourth fundamental circuit element and have been the focus of intense research since the theory~\cite{14,84} was recently united with a practical demonstration~\cite{15}. As they can hold a state, they may be useful as computer memory~\cite{13} and because synapses~\cite{84} can be described by the memristor theory, it is anticipated that memristor devices could operate as artificial synapses in neuromorphic computing paradigms~\cite{11} and there are several theoretical studies~\cite{71,41,214,215,216} that suggest that this approach may be fruitful and ongoing experimental projects~\cite{102,117} (including ours) to test it out. 

The definition of the memristor is a non-linear relationship between voltage and current, due to a linear relationship between charge and flux, where the memristor resistance is a function of either charge or flux~\cite{14}. In practical tests, the memristor is best identified by three `fingerprints': such as the existence of a pinched hysteresis curve (specifically a Lissajous curve)~\cite{276}. 

In terms of technological approaches and types of materials, memristors are very similar to Resistive Random Access Memory, ReRAM (also known as RRAM). The relation between ReRAM and memristor technologies is not a settled one: it has been suggested that all resistance switching memories (include ReRAM) are memristors~\cite{119}, that memristor theory can be only used to describe parts of the operation of ReRAM~\cite{WasersBook} and that some ReRAM requires an extension to memristor theory~\cite{254}. 

ReRAM devices are known to exhibit a rich set of behaviours including bipolar switching, BPS, a slow changing of resistance and unipolar switching, UPS, a more sudden switching regime~\cite{155}. To test for BPS, devices are subjected to a change in voltage and polarity of voltage, to test for UPS the devices are usually only tested with one voltage polarity. The results of subjecting UPS switching devices to a switch in polarity as well as voltage magnitude is not generally published and informal discussions with ReRAM scientist suggest that it's not generally tested.

Titanium dioxide is not the only transition metal oxide that can switch via both BPS and UPS, a few examples are SrTiO$_3$~\cite{169,170}, NiO~\cite{164}, CuO~\cite{162}, ZnO~\cite{160}, MnO$_x$~\cite{176}, and both binary and perovskite oxides are capable of it. The exact mechanisms are not completely understood, even if we restrict our focus to TiO$_2$. The UPS switching has credited to: Joule heating~\cite{155}, metal-semiconductor transitions~\cite{155}, anion migration~\cite{155}, crystalline TiO$_2$-amorphous TiO$_2$ phase transition via conduction heating and breaking~\cite{159}, conical conducting filaments made of a Magn\'{e}li phase~\cite{165}, fractal conducting filaments~\cite{186}, raising and lowering Schottky barriers via bulk transport of oxygen~\cite{169}, and conductance heating causing lateral transport of conducting filaments~\cite{168}. Similarly, BPS switching has been credited to: Joule heating~\cite{155}, metal-semiconductor transitions~\cite{155}, anion migration~\cite{155}, Magn\'{e}li-insulator phase transitions~\cite{136}, Ti$_4$O$_7$ conducting channels~\cite{154}, raising and lowering Schottky barriers via interface-based transport of oxygen~\cite{169} and migration-caused slow breaking of conduction filaments~\cite{168}. There is also a debate as to what the structure of the less-conducting thin-film TiO$_2$ is, with rutile ($r$-TiO$_2$)~\cite{159} and amorphous titanium dixoide ($a$-TiO$_2$) being the most popular suggestions. Note, that $a$-TiO$_2$ is also suggested as the conducting form of TiO$_2$ and that Magn\'{e}li phases are sheer planes from $r$-TiO$_2$. However, the ReRAM community seems to have settled on the slow drift of vacancies or ions as being the root cause of BPS. This type of switching resembles the theoretical memristor curves and is the same mechanism as that put forward in a forthcoming theoretical paper~\cite{F1}. The unipolar switching mechanism is most likely related to the creation of conducting filaments of doped material~\cite{155,130,159,168,186} and 
can be described by an extension to memristor theory~\cite{254}. This overlap in terms of functionality and theoretical viewpoint strongly suggest that ReRAM research is highly relevant to memristor research, a view which has been also taken up by some memristor researchers~\cite{197}, thus we shall not limit ourselves to the memristor literature. 

A recent paper~\cite{28} describing the creation of a solution processed TiO$_2$ sol-gel memristor that could be fabricated on a flexible plastic substrate has attracted much interest because it could lead the way to easier fabrication of memristors as well as the possibility of flexible or wearable memristor circuitry. The archetypal memristor~\cite{15} was made with platinum electrodes which are expected to have no effect on the TiO$_2$ switching, whereas the sol-gel memristor had aluminium electrodes~\cite{28}. There have been several examples in the ReRAM literature of aluminium oxide playing an essential role in resistance switching mechanisms~\cite{177,178}, including devices based on TiO$_2$ and aluminium~\cite{174,159,189,123}, the addition of Al$_2$O$_3$ to improve switching~\cite{173}, Al-Al$_2$O$_3$ resistive switching~\cite{172,179} and even the fabrication of Al-Al$_2$O$_3$-Al based flexible switching memory~\cite{171}. Despite this weight of expectation that the aluminium electrodes might 
effect the switching in the flexible memristor, the authors did not report detailed tests and instead simply stated that the `switching behaviour in our devices cannot be attributed to the aluminium (or aluminium oxide), since it was also observed from devices with noble metal (gold, silver and platinum contacts'. It is known that TiO$_2$ can switch by itself (see for example~\cite{136,15}), but it is important to know if the aluminium electrodes are having an effect, as this would make the flexible memristor not a simpler-to-manufacture version of the Strukov memristor~\cite{15} but instead a more complex system. It is suggested that a more complex system may well have a richer set of behaviour and thus a wider range of uses. 

In this paper, we have undertaken an investigation of the effect of electrode material on the TiO$_2$ sol-gel memristor and in so doing we have discovered a much richer set of possible behaviours than has been reported previously and made an attempt to elucidate how to control which behaviour is selected. It is envisaged that this level of understanding will allow researchers to more finely control the aspects of their memristor devices and allow creation of useful and practical memristor devices.


%

\section{Methodology}

The standard memristors were prepared by first sputtering aluminium onto a plastic substrate via a mask. The sol-gel preparation is based on that found in~\cite{28,96} and is described in detail elsewhere~\cite{M0}. The final preparation was diluted 1:10 in dried methanol to make the spin-coat solution. 3ml of spin-coat solution was spun at 33 r/s for 60s and left for at least an hour (the time taken for the sol to convert to the gel~\cite{95}) before the second set of electrodes were sputtered on top. The electrodes were 4mm wide and crossed at 90 degrees giving a 16mm$^{2}$ active area. Profilometer tests revealed that the TiO$_2$ gel layer was approximately 44nm ($\pm2.5nm$) thick. 

6 batches of devices were fabricated and in total 130 devices were tested. To test the fabrication methods devices were left in clean room air to hydrolyse (R series) or put into vacuum (V series) or the top electrode size was changed to 1, 2, 3, 4 or 5mm, giving active areas of 4, 8, 12, 16 or 20mm$^{2}$ (D series). The D series memristors were hydrolysed under vacuum. Comparison devices were made with: A, gold electrodes top and bottom (Au-TiO$_2$-Au memristors); B, aluminium bottom electrodes and gold top electrode (Au-TiO$_2$-Al memristors); and C, the standard aluminium top and bottom electrode (Al-TiO$_2$-Al), all these devices were made with 4mm wide electrodes. From the memristor literature it is expected that the Au-TiO$_2$-Au devices should switch similarly to the Al-TiO$_2$-Al memristors, from the ReRAM literature it is expected that the aluminium oxide has a role to play in the switching.

Virgin memristor devices were used for all comparisons and were measured with linear I-V sweeps on a Keithley 2400 sourcemeter between a range of $\pm3.5V$ which is the same range as ~\cite{28}, with a voltage step size of 0.05V and a dwell time of 2s: this is the D.C. equivalent of an A.C. voltage of 1mHz, a frequency that it too slow to source any other way from the sourcemeter. Twelve memristors were tested in each case, the source was always connected to the bottom electrode to ensure comparability. For the Al-TiO$_2$-Au memristor virgin runs, six devices were connected one way (Earth to Au, source to Al) and six the other way. For these devices a second run done connecting the wires the opposite way round (to reverse the asymmetry). The compliance current was 2mA unless stated.

As in~\cite{28} our electrodes were sputter coated, the TiO$_4$\textit{(sol)} is then spun on-top and hydrolyses to the gel~\cite{98}: 

\begin{quote}
 Ti(OH)$_{4}(sol) \rightarrow{\mathrm{H}_{2}\mathrm{O}}$ TiO$_{2}(gel)$
\end{quote}

before the next set of electrodes are sputtered on top. The Al-TiO$_2$-Al were prepared in two different ways, depending on whether they were left in the clean room atmosphere to hydrolyse or kept under vacuum. Firstly, aluminium oxide forms very quickly and secondly the TiO$_2$ may oxidize the bottom electrode, and so we expect the vacuum prepared memristors to have an Al$_2$O$_3$ layer, however it may be a different thickness or the TiO$_2$ layer may have a different form under the two different preparation methods.

Photographs were taken with a Canon Powershot G6 using frontal and side-light from a tungsten desk lamp using a sheet of white paper as a soft-box. Micrographs were taken using on a 3D Hirox digital microscope using both frontal and side-lighting.

\section{Results and Discussion}

\subsection{Types of Memristors Tested}

\subsubsection{Au-TiO$_2$-Au Memristors}

\begin{figure}[t!]
 \centering
 \includegraphics[width=3.5in]{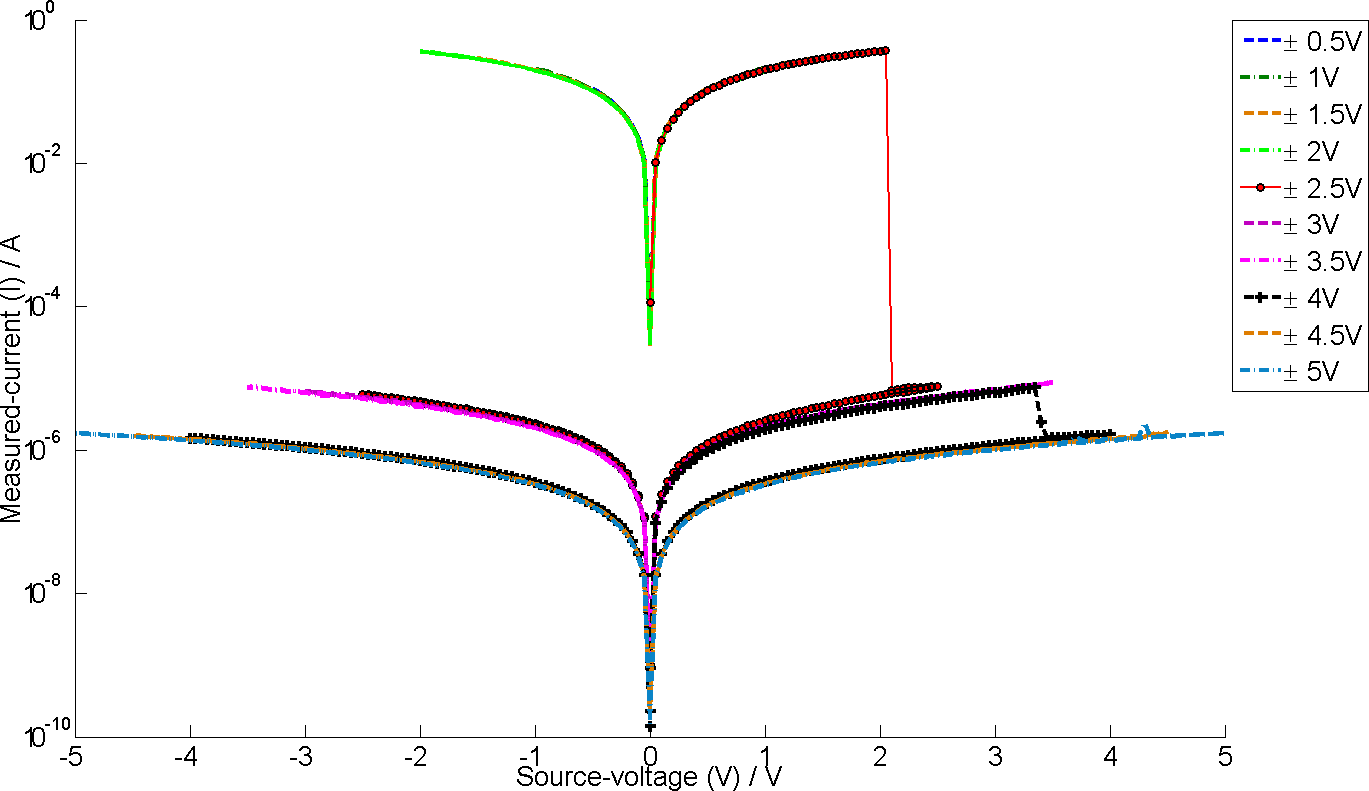}
 \caption{An example of an Au-TiO$_2$-Au device. Repeated I-V curves were run with increasing voltage ranges. The device starts off in a low resistance state and switches down 5 orders of magnitude (during the $\pm2.5$V I-V curve) at just over 2V. The device can switch out of this state if subsequently taken above 3V.}
 \label{fig:A4}
\end{figure}

\begin{figure}[htbp]
 \centering
 \includegraphics[scale=0.75,keepaspectratio=true]{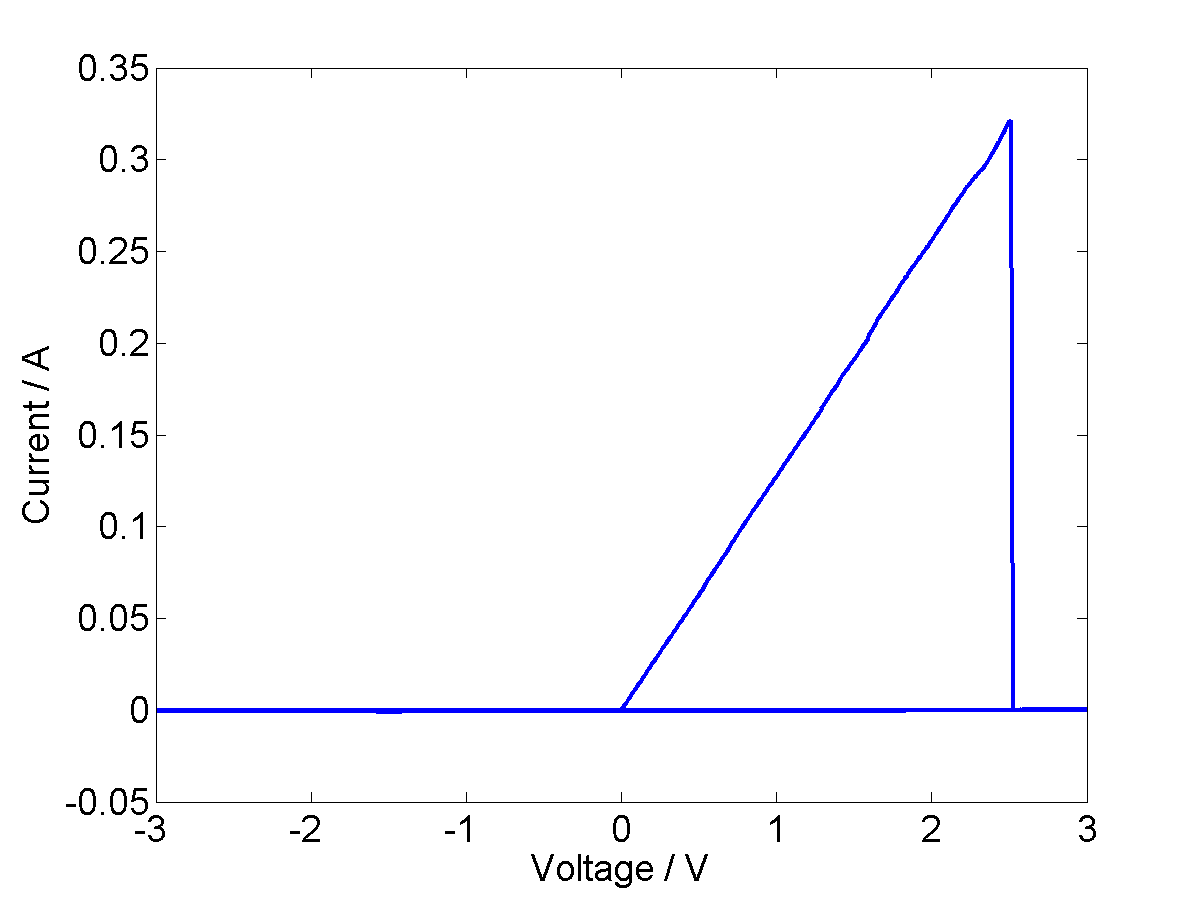}
 \caption{An example of a typical Au-TiO$_2$-Au virgin run profile. These devices were much more similar in that all the virgin runs were classified as the same type. The LRS is a straight-line ohmic conduction, this then breaks given a HRS which is 2-3 orders of magnitude smaller. Note that the LRS current is an order of magnitude higher than the UPS in the Al-TiO$_2$-Al memristors.}
 \label{fig:AuAuExample}
\end{figure}

\begin{figure}[htbp]
 \centering
 \includegraphics[scale=0.75,keepaspectratio=true]{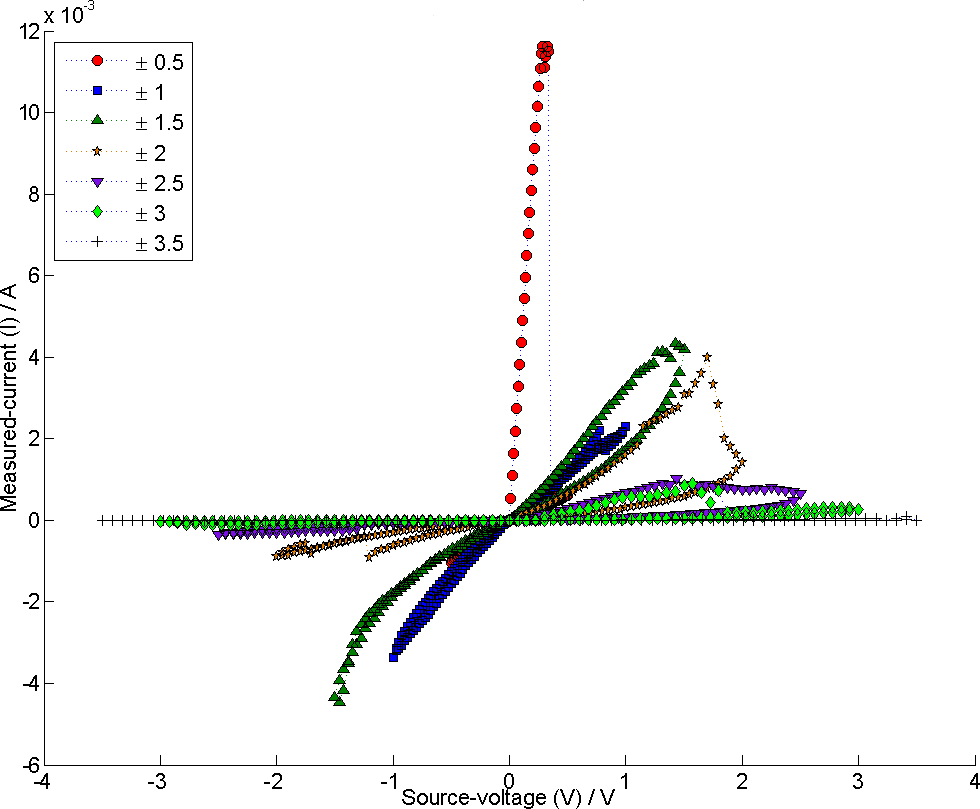}
 \caption{An example of repeated runs over a changing voltage range. Once the initial ohmic conduction pathway breaks, the device exhibits high current memristive behaviour.}
 \label{fig:At7}
\end{figure}

Using the gold electrodes gave a fuse-like behaviour, see figure~\ref{fig:Combo}A. The devices start off in a very low ohmic resistance state, see figure~\ref{fig:AuAuExample}, with $R_{1}\sim4.6\Omega$. This state breaks down at just over 2V to a higher resistance state, $R_{2}\sim3.2\times10^{5}\Omega$, and secondary breakdown is observed between 3 and 4V to $R_{3}\sim2.9\times10^{6}\Omega$, as shown in figure~\ref{fig:A4}. This happened in all tested devices and was not reversible over a short timescale. Reversing the order that the device went round the I-V curve meant the switching behaviour was observed at a negative voltage. These devices might have applications in Write Once Read Many (times) WORM memory as the separation in current between $R_1$ and $R_2$ is in the order of 5 orders of magnitude, see figure~\ref{fig:A4}. The Low Resistance State, LRS, and High Resistance State, HRS, are generally both linear and thus these devices are not memristors (they are ohmic resistance switches). We suspect that disruption of gold bridges causes the observed fall in resistance, as in~\cite{188}. A few devices showed memristive behaviour after the first fuse breaking, see for example~\ref{fig:At7}, although it was not very stable, which we expect is the action of TiO$_2$ gel.

To test for unipolarity, a device was run backwards, i.e. the negative half of the curve was run first. As would be expected if the size rather than the polarity the voltage caused the effect, the device switched on the negative side first. However, it switched at $\approx-4V$ rather than $\approx 2V$, suggesting that the devices are not symmetric. 

The mechanism of these devices is unknown. Silver Chalcogenide memristors operate via the reproducible formation and breaking of silver nanowires with the chalcogenide~\cite{104}. Gold is known to form nanowires which are highly extrudable (and can be used in break junctions). Filamentary conduction states have been recorded in thin continuous gold films on insulating surfaces~\cite{188}. Given these facts, and the fact that the Au-TiO$_2$-Au device currents are much higher than the Al-TiO$_2$-Al device currents, it could be that metallic gold filaments are forming, passing through the very thin TiO$_2$ layer and forming a short-circuit. These filaments could then be being broken by Joule heating.

\subsubsection{Al-TiO$_2$-Al}

One advantage that the sol-gel based fabrication method has over sputter or atomic deposition of TiO$_2$ is that the sol-gel memristors can switch during their virgin I-V run and do not need a forming step. Most reported ReRAM devices require a forming step because deposited TiO$_2$ has a very regular and crystalline structure, therefore holes or defects must be introduced~\cite{154} into the material (N.B, another approach is to add oxygen chemically during fabrication~\cite{143}) to cause appreciable conduction. The forming process generally involves pulsing~\cite{154} or increasing the voltage to high levels with a low compliance current~\cite{155,169}, this may change the device by moving the oxygen anions~\cite{154} or may change the structure via Joule heating~\cite{155} and furthermore both those statements may be equivalent in some devices, i.e., they are equivalent if Joule heating causes a structural rearrangement which moves the oxygen ions and not equivalent if the oxygen ions drift without significant heating. Forming is not necessary in the memristors reported here as the gel-form TiO$_2$ structure is much more amorphous and already contains defects and thus these memristors can switch right away.

\paragraph{Behaviour and Device Properties}

\begin{figure}[!tbp]
\centering
\subfigure[]{\includegraphics[width=0.49\textwidth]{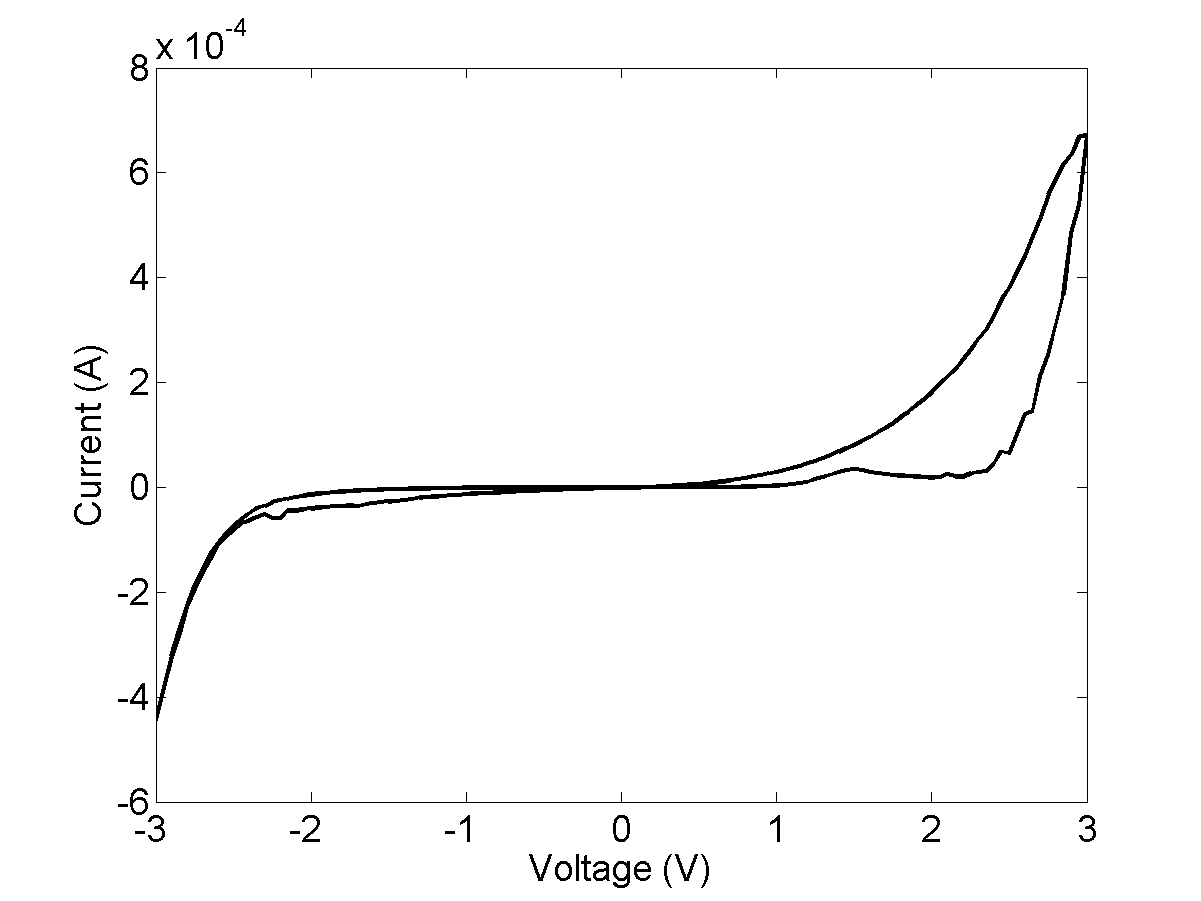}}
\subfigure[]{\includegraphics[width=0.49\textwidth]{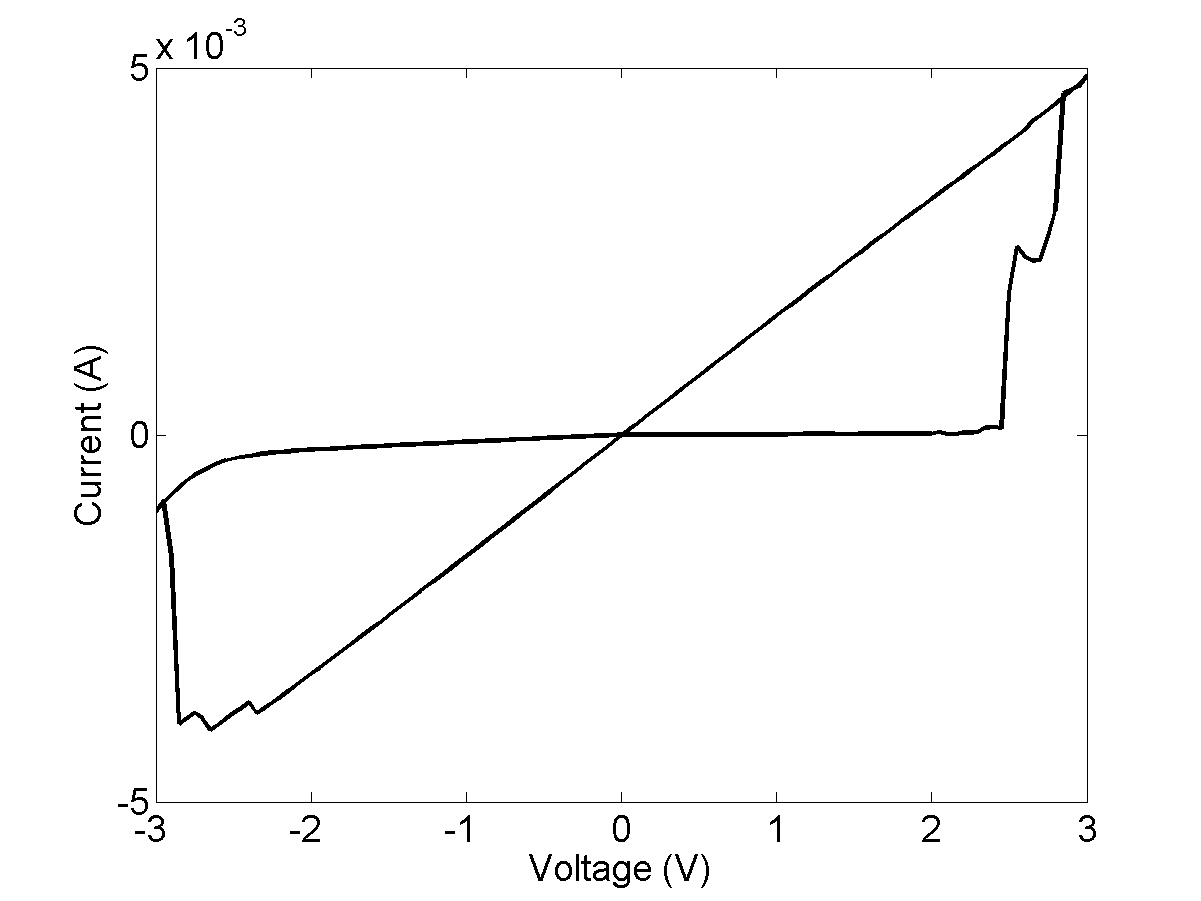}}
\caption{Different behaviours seen in Al-TiO$_2$-Al: a: Type A behaviour (curved); b. Type B behaviour (triangular)}
\label{fig:Combo}
\end{figure}

Changing both electrodes to aluminium increases the richness of the behaviour compared to gold electrode devices and improves the repeatability and stability of behaviour. We observed two types of behaviour, curved, see figure~\ref{fig:Combo}a and triangular, see figure~\ref{fig:Combo}b, I-V profiles. The curved switching profile is integrable (i.e. there are no discontinuities). These devices are very similar in form to the theoretical memristor curves~\cite{14} in that they are pinched, curved hysteresis loops, they differ in that they are more pinched close to the origin than at higher voltages. This may imply the existence of a threashold for memristive behaviour, which makes sense chemically and allows the devices to be `read' at low voltages with very little or zero effect on the resistance value. Devices exhibiting this sort of switching have a HRS and LRS within the same order of magnitude which are usually in the range $10^{-6}$A$-10^{-3}$A, most often around $10^{-4}$A, see figure~\ref{fig:Combo}a. We suggest that these devices closely resemble ReRAM BPS switching (as in~\cite{155}).

The triangular switching is shown in figure~\ref{fig:Combo}b. The devices start off in the HRS and then switch over a very small voltage range (i.e. there is a discontinuity in the current-time profiles) to the LRS which is orders of magnitude smaller and ohmic. The LRS gives currents in the 10$^{-3}$A range and these devices never reach the 10$^{-1}$ current range seen in the gold electrode devices. The HRS is curved and resembles the curved switching devices. Thus, we think the LRS is caused by filaments that connect after the device has been under a large enough voltage, and when not connected the device shows bulk memristance similar to that seen in the curved devices. These filaments give an ohmic response and could be considered as fuses. Changing the compliance current can allow a ReRAM device to switch between UPS and BPS behaviour~\cite{157}, and thus we tested if changing the compliance current in these sol-gel devices would cause a transition in behaviour; it did not. 

The sol-gel memristors are capable of both BPS-like and UPS-like switching. The coexistence of BPS and UPS switching has been observed before in thin-film TiO$_2$ before~\cite{183,159,157}, although none of these were sol-gels, and in SrTiO$_3$ sol-gel based ReRAM~\cite{169}. The triangular-switching resembles that reported by Gergel-Hackett et al in their paper on sol-gel TiO$_2$ memristors~\cite{28}. Our devices go round the I-V loop backwards compared to the Strukov et al memristor~\cite{15}, as they start in the high conducting state and switch off, rather than the other way round. This makes sense if the gel phase is or is largely $a$-TiO$_2$ as it will have a high conductivity to start with, whereas the Strukov memristor is closer to a crystalline form of TiO$_2$. Previous work using the same sol-gel solution with drop-coated devices~\cite{M0} showed that the titanium dioxide gel layer was mostly the amorphous form, $a$-TiO$_2$. 

However, our triangular switching is not necessarily UPS. Unipolar devices can be SET and RESET with only a single polarity voltage and compliance current control (as described in~\cite{157}), however this process was not found to reset these devices. A RESET to a negative voltage loop was, i.e. they are bipolar switches. 

After the triangular-switching devices have been taken to a high voltage ($>\pm3.5$), in subsequent I-V measurements the triangular-switching behaviour would be replaced by the curved behaviour. Once this had happened the devices were not properly reset by a negative voltage and were less reproducible: further repeated trips around the I-V curve showed a continuous memory effect whereby the HRS of one run would become the LRS of the next. This is consistent with the filament breaking leaving the memristor only operating via the bulk drift of vacancies.

\begin{figure}[!tbp]
\centering
\subfigure[]{\includegraphics[width=0.49\textwidth]{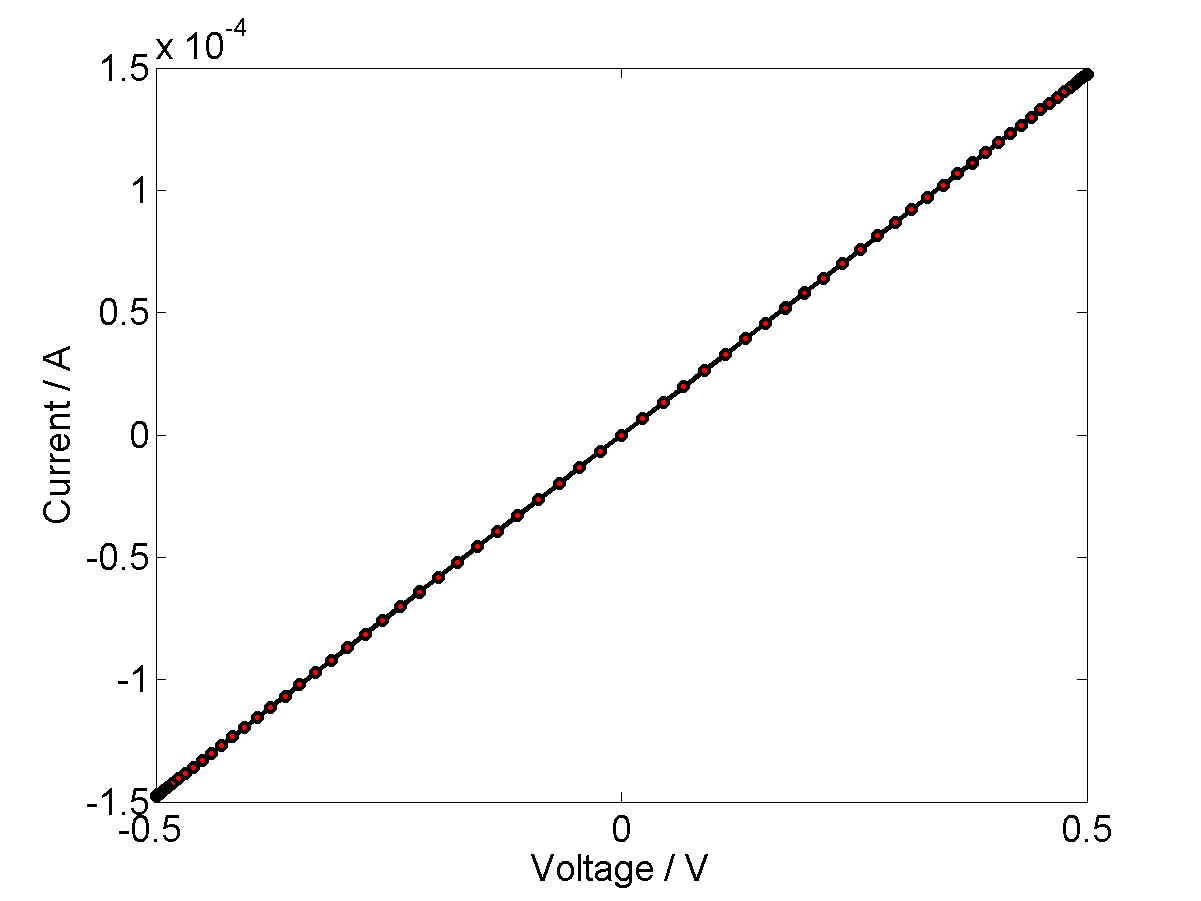}}
\subfigure[]{\includegraphics[width=0.49\textwidth]{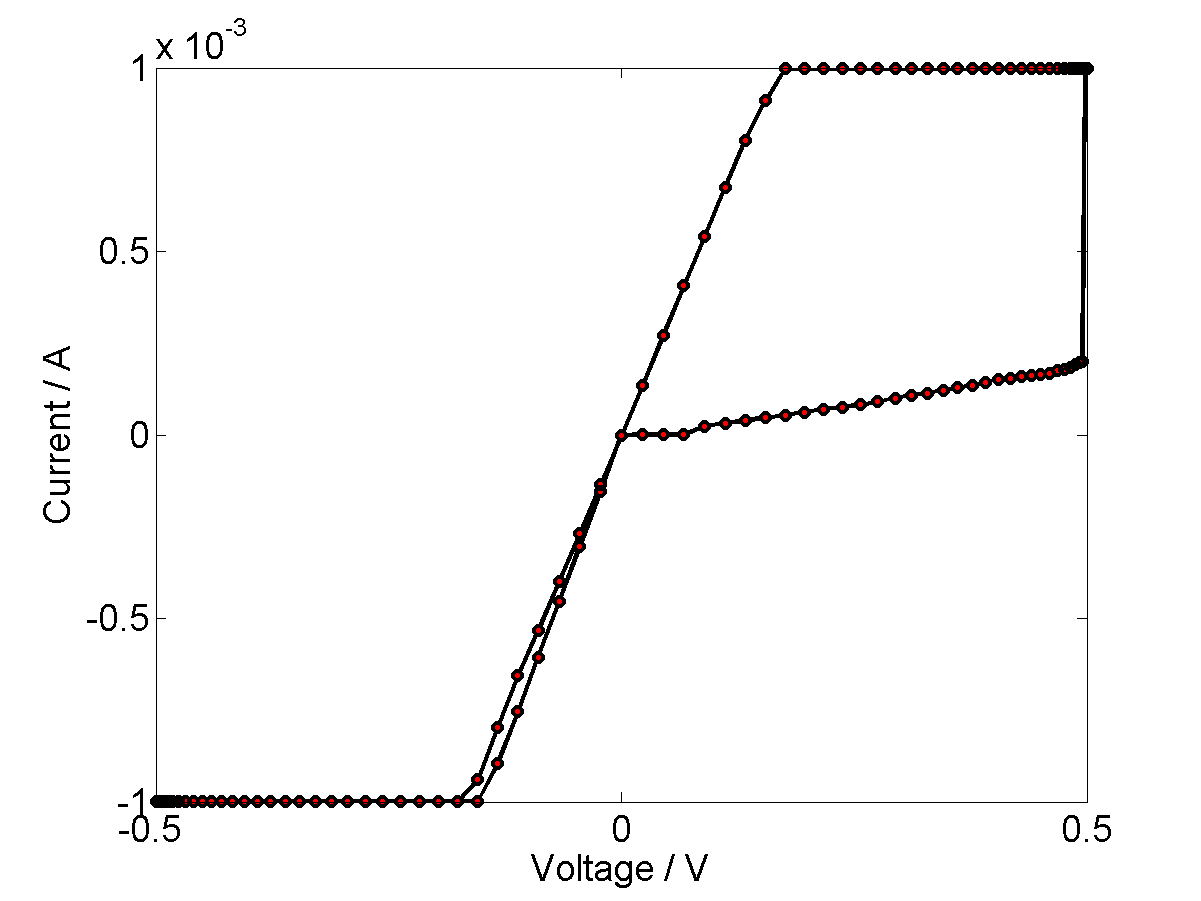}}
\subfigure[]{\includegraphics[width=0.49\textwidth]{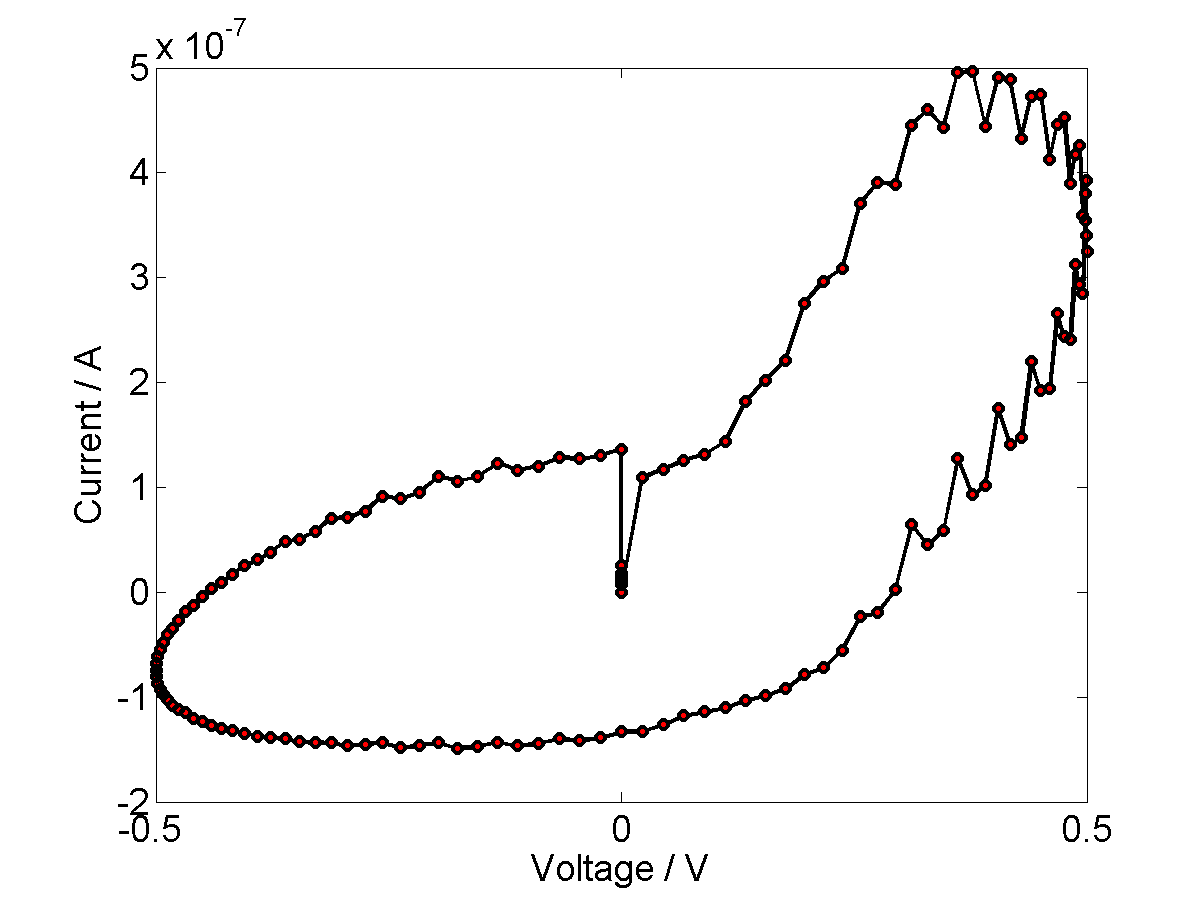}}
\subfigure[]{\includegraphics[width=0.49\textwidth]{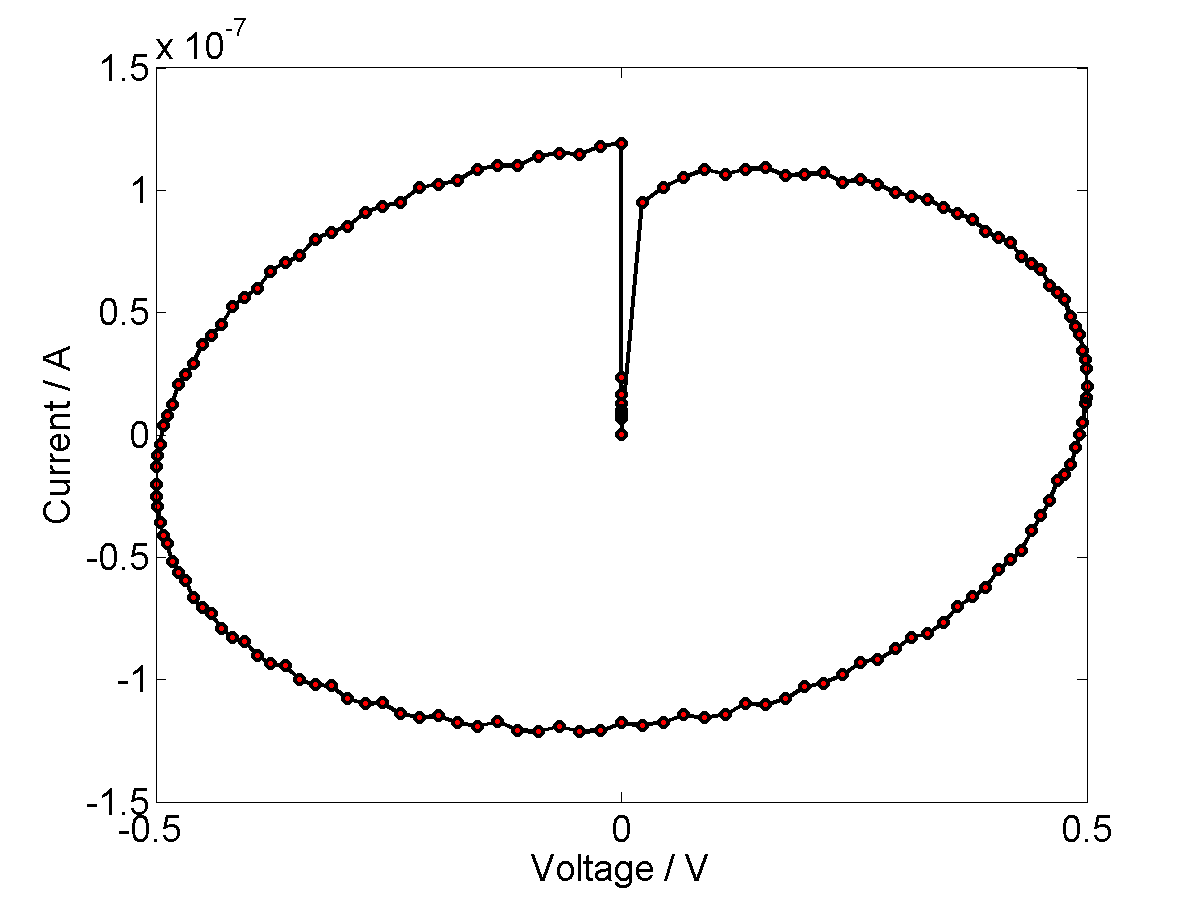}}
\subfigure[]{\includegraphics[width=0.49\textwidth]{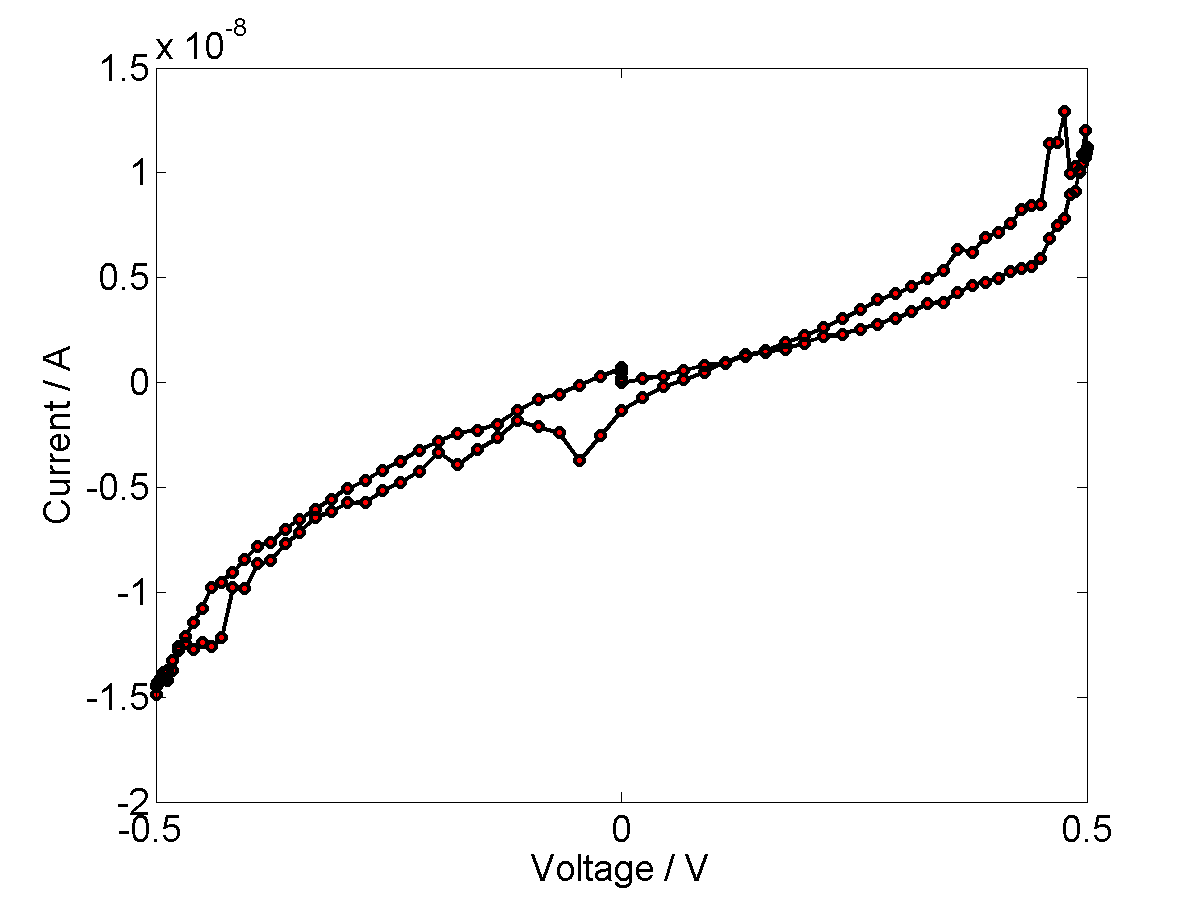}}
\subfigure[]{\includegraphics[width=0.49\textwidth]{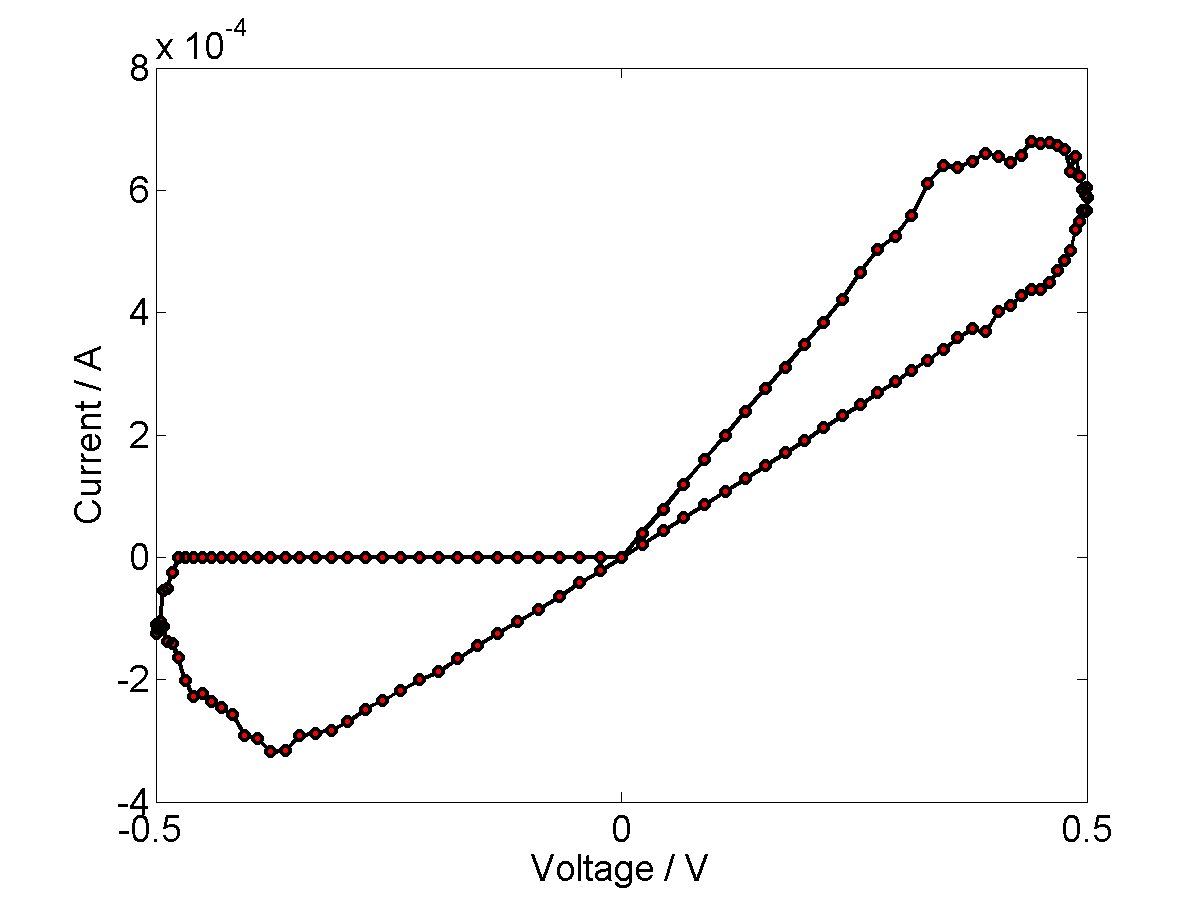}}
\caption{Different classification of device behaviour: a. Ohmic; b. Switch-Ohmic; c. Jelly Bean; d. Circle; e. Nearly UPS; f. Nearly BPS.}
\label{fig:Types}
\end{figure}

It was found that measuring an I-V curve over $\pm0.5V$ on virgin devices would show two types of behaviour, ohmic and a distinctive open curve as shown in figure~\ref{fig:Types}e and previously reported for TiO$_2$ thin films~\cite{183}. Devices which were ohmic over this range had starting resistances between 26\Ohm~ to 352\Ohm~ tended to exhibit triangular behaviour over higher voltage ranges, those which were curved tended to exhibit curved over a higher voltage range, providing quick device classification. 

Similar devices fabricated by drop-coating the same sol-gel solution onto sputter-coated Al contacts have been tested over the same range~\cite{M0} and only displayed curved switching over this voltage range. This suggests that the thickness of the layer has an effect on the presence or absence of triangular behaviour (over a given voltage range), which may suggest that the ohmic state in the triangular switching was due to the formation of filaments and the curved I-V and LRS of the triangular switching was due to unconnected filaments or background drift of oxygen vacancies. 

\paragraph{Device Structure}

\begin{figure}[htbp]
 \centering
 \includegraphics[scale=0.5,keepaspectratio=true]{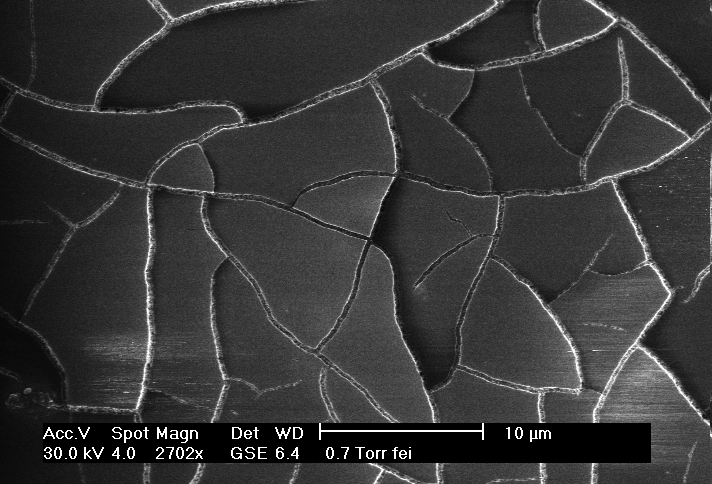}
 \caption{The surface of spun-coated TiO$_2$(gel) phase. It resembles paddy fields. The composition of the boundaries is not known, however they are not believed to be cracks in the material.}
 \label{fig:FeildOfSurface}
\end{figure}

\begin{figure}[htbp]
 \centering
 \includegraphics[scale=0.5,keepaspectratio=true]{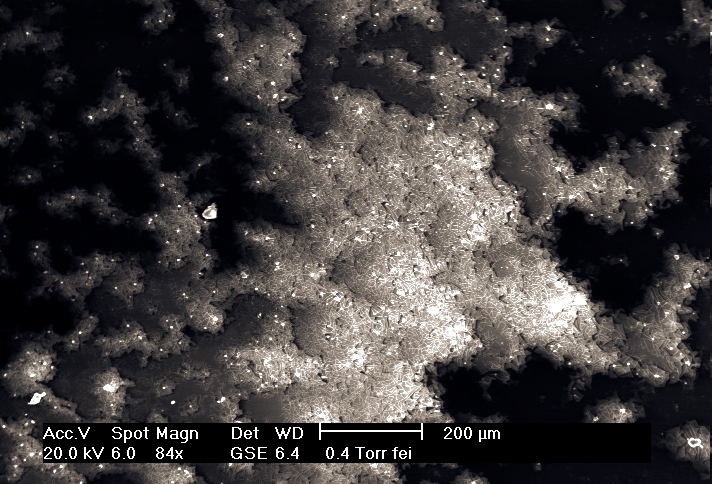}
 \caption{A zoomed out view of the spun-coated TiO$_2$ (gel) surface, showing the possible height difference in the paddy-field like structure.}
 \label{fig:Wide}
\end{figure}

A part of the plastic covered with only TiO$_2$ gel layer was imaged with a scanning electron microscope. Figure~\ref{fig:FeildOfSurface} shows the paddy-field-like structure of the surface. The `cracks' were not a result of electrode beam breakage as they didn't increase whilst the material was imaged. They could be a result of the gold sputtering required to image the surface or it could be that the flat planes are relatively regular islands of TiO$_2$ and the `cracks' are grain boundaries, the formation of extended defects along grain boundaries has been seen before~\cite{218}. Note, figure~\ref{fig:Wide} shows a much wider view and we can see that the structure is like paddy-fields in that there seems to be a large difference in brightness across the sample, as the difference in brightness in a SEM micrograph is a convolution of the distance from the electron microscope tip and the electronic conductivity of the region, thus it could well be that the brighter areas are areas of more conductive TiO$_2$ and as we know some parts of the material are in this form, we can't rule out that they would be laterally distributed as well as distributed along the line of electron movement. Devices were sliced in half and imaged by transmission electron microscopy (TEM) and do not show obvious filaments, however, given the size of the active area and the difficulty of slicing the device more than once, absence of filaments in these micrographs does not necessarily prove their non-existence.

\begin{figure}[htbp]
 \centering
 \includegraphics[scale=0.75,keepaspectratio=true]{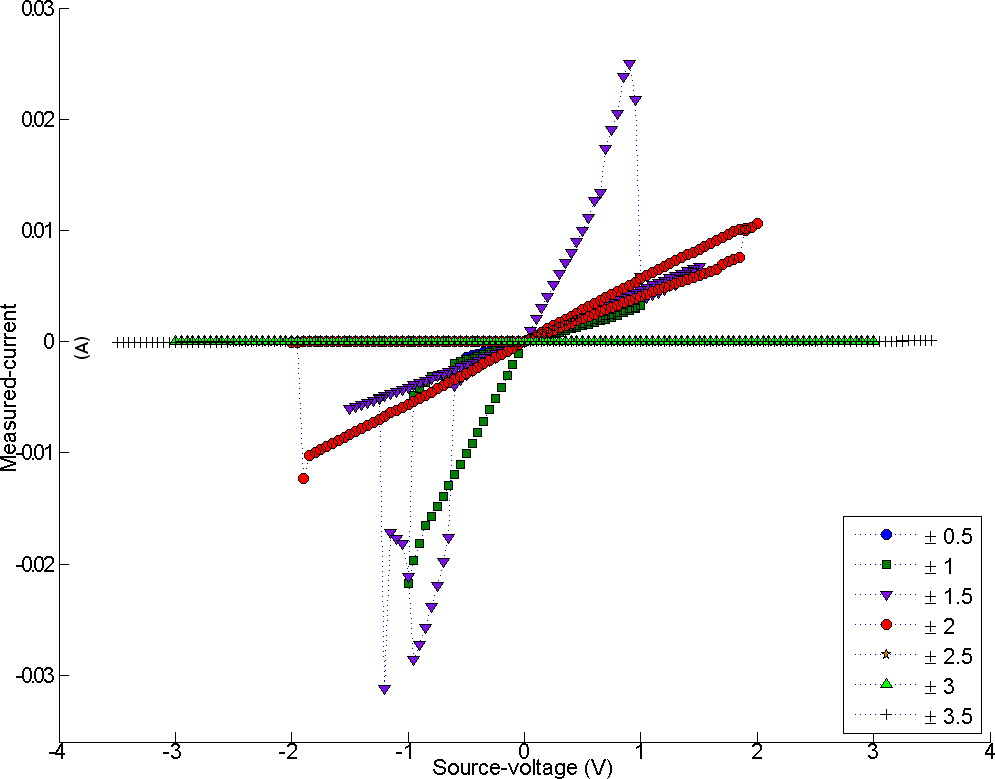}
 \caption{A single device was run repeated times over a gradually increasing range. The classification run at $\pm0.5V$ was a straight line, suggested UPS behaviour. We see UPS behaviour between $\pm1 - \pm2V$, after that, the device is BPS. Note that as the switching effect can happen in either the positive or negative part of the I-V curve, it may be classifiable as true UPS.}
 \label{fig:R42UPSRepeatExperiment1}
\end{figure}

\begin{figure}[htbp]
 \centering
 \includegraphics[scale=0.75,keepaspectratio=true]{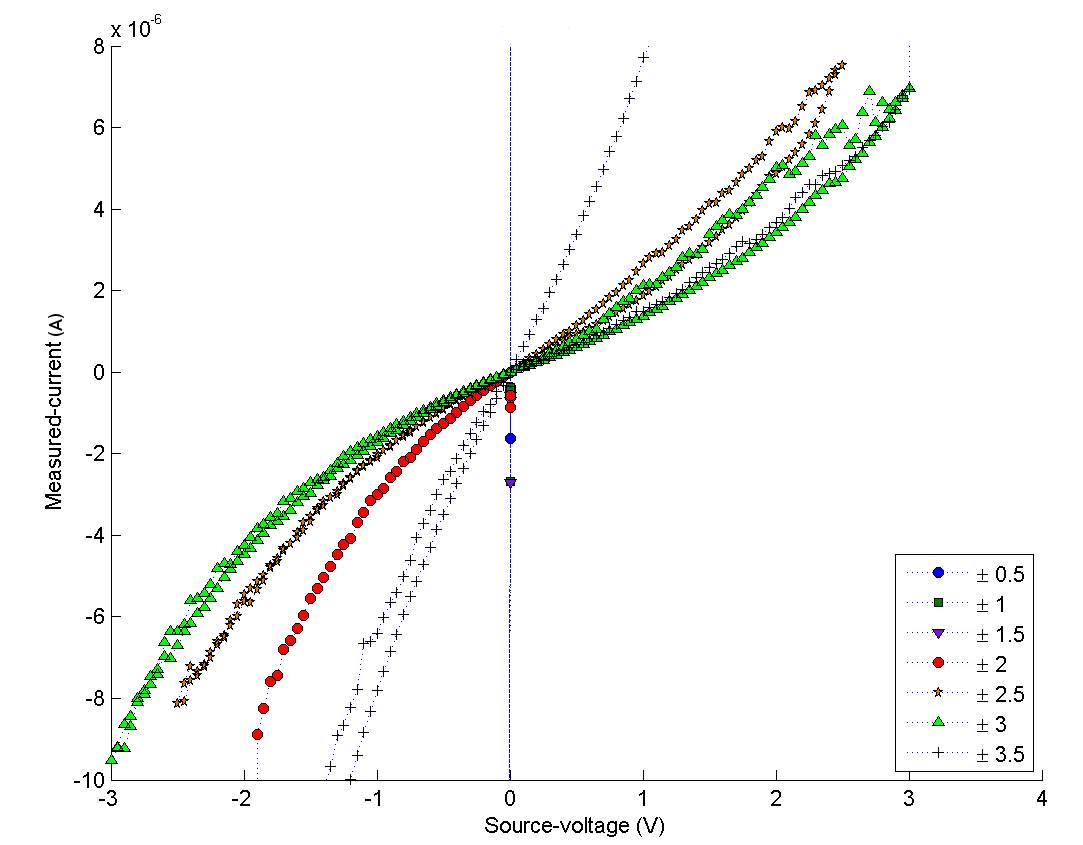}
 \caption{A close up of figure~\ref{fig:R42UPSRepeatExperiment1} showing the BPS behaviour of both the higher voltage range runs and the run at $\pm2V$ after the UPS switching has stopped. This may indicate that the material is capable of, or maybe even undergoing, BPS while doing UPS. This fits with the suggestion that UPS is related to filaments forming and BPS is related to a `bulk' phase change.}
 \label{fig:R42UPSRepeatExperiment2}
\end{figure}

\paragraph{The Effect of Atmospheric Conditions During the TiO$_2$ gel Formation on the Final Device Characteristics}

The TiO$_{2} (gel)$ layer was formed under two different atmospheric conditions: under vacuum overnight (V series memristors) and exposed to clean-room air overnight (R series memristors). Air humidity has been noticed to effect the qualities of memristor-like ReRAM in the past~\cite{175}. although this has not been reported with TiO$_2$ devices. The V series were only exposed to air for a short time during transfer to from the sputterer to vacuum. The air that both memristors were exposed to was clean room air, which would have a lower proportion of water and dust than standard laboratory air. Both sets of memristors were left in their respective places overnight, giving the gel plenty of time to react. 

A comparison between the two types is shown in table~\ref{tab:VVsR}, they were tested with two different values of compliance current (c.c). It was thought that the compliance current, 2mA, would be high enough to not limit the behavioural choices of the device (as discussed earlier), however an even higher compliance current (1A), which is effectively unlimited as it is higher than any current drawn by these devices over this range, was found to change some of the behaviour. Note, that these data are four different sets of 12 memristors undergoing their virgin run, which is enough to give an indication but not enough to be statistically reliable, so the conclusions below are tentative. 

In addition to curved (A) and triangular (B) designations, four others were added to the classification in table~\ref{tab:VVsR}: `Mixed' which were I-V curves that seemed to switch from triangular to curved type behaviour; `half-triangle' which is a triangle in the positive I-V part of the graph qualitatively similar to that seen for Au-TiO$_2$-Au as shown in figure~\ref{fig:At7} but with a current an order of magnitude less; `linear' which an ohmic resistor I-V profile and `unconnected' which is a broken device (equivalent to an open circuit). Curved and triangular types are useful as memristor devices, linear, half-triangle and mixed are only useful if they act like curved or triangular after their virgin runs (which some, but not all, do). As mixed I-V curves are only seen with the high compliance current and the sum of the number of triangular and mixed I-V curves at high c.c. is comparable to the number of triangular I-V curves at low c.c., $\frac{7}{12}:\frac{8}{12}$(V), $\frac{7}{12}:\frac{5}{12}$(R), we conclude that the low compliance current preserves the triangular I-V behaviour by not allowing it to break to become mixed behaviour. Similarly, from the presence of linear I-V curves at a low c.c. ($\frac{2}{12}$(V) and $\frac{4}{12}$(R)), a non-presence of them at the high c.c. and the presence of half-triangle I-V curves ($\frac{2}{12}$(V) and $\frac{3}{12}$(R)) and the similar numbers of these categories, we think that the low compliance current prevents the filaments from breaking, phenomenon which gives rise to the half-triangle designation at high compliance currents. The lower number of triangular devices with left in clean room air (R) compared to those left in vacuum at both compliance currents (5:8 and 3:5) indicates that the vacuum treatment yields much more useful devices. The number of devices exhibiting curved behaviour is not affected by either the two treatments during gel formation or the compliance current and is approximately a sixth of the tested devices.

\begin{table}
\begin{tabular}{|c|c|c|c|c|}
\hline
Types of memristors $\: \rightarrow$ 	& \multicolumn{2}{|c|}{V memristors} & \multicolumn{2}{|c|}{R memristors} \\
\hline
Compliance current $\: \rightarrow$ 	& $2mA$	& $1A$ 		& $2mA$	& $1A$	\\
\hline
Memristor types seen $\: \downarrow$	& \multicolumn{4}{|c|}{   } \\
\hline
Curved (Type A)				& 2	& 2		& 3	& 2	\\
\hline
Triangular (Type B)		& 8	& 5		& 5	& 3	\\
\hline
Mixed									& $-$	& 2		& $-$	& 4	\\
\hline
Half Triangle					& $-$	& 2		& $-$	& 3	\\
\hline 
Linear								& 2	& $-$		& 4	& $-$	\\
\hline 
Not connected					& $-$	& 1		& $-$	& $-$	\\
\hline
\hline
Total 								& 12	& 12		& 12	& 12	\\
\hline
\end{tabular}

\caption{The effect of atmospheric conditions during TiO$_2$ gel formation on the final device characteristics during a virgin run. The V type were put under vacuum overnight before sputtering on the second set of the electrodes, the R type were left in the clean room air to hydrolyse. Note, Linear is an ohmic resistor over the range tested.}
 \label{tab:VVsR}
\end{table}

\paragraph{The Effect of device size}

64 devices were tested and 7 different device behaviours were observed with the following characteristics: 1. Ohmic: Devices were ohmic over the measurement range ($\pm$0.5V) ; 2. Switch-Ohmic: Devices were ohmic over the measurement range, but switched from one ohmic value to another; 3. Jelly-bean: Devices were a pinched but open curve; 4. Circle: Devices were an open circle ; 5. Nearly UPS: Devices switched over the range but were not strictly ohmic; 6. Nearly BPS: Devices showed a pinched hysteresis loop; 7. Not connected: Devices were not connected over the range. 

\begin{table}[htpb]
	\begin{tabular}{|c|l|r|r|r|}
\hline
	Type No.& Type 			& No. of devices & Average Order 	& $\sigma$ 	\\
		& 			&		& of magnitude		& 		\\
\hline
	1 	& Ohmic 		& 24		& 3.24			& 1.27	\\			
	2	& Switch Ohmic		&  9		& 3.78			& 0.83	\\
	3	& Jelly Bean		& 16		& 7.19			& 0.40	\\
	4	& Circle		&  3		& 7.67			& 0.58	\\
	5	& Nearly UPS		&  4		& 4.00			& 0.82	\\
	6	& Nearly BPS		&  5		& 8.80			& 0.45	\\
	7	& Unconnected		&  3		& 10.67			& 0.57	\\
\hline

	\end{tabular}
	\label{tab:DiffElecTypes}
	\caption{Different device classifications have different current ranges. Note that due to the compliance current, the order of current was not allowed to exceed 3.}
\end{table}

The Nearly UPS and Nearly BPS device designations were generally low-quality devices that switched over such a low voltage range. It was found with previous batches of devices (namely those described in the previous section) that devices which were Ohmic over $\pm$0.5V would display triangular behaviour over larger voltage ranges, and those which had open curves over this range would display curved I-V curves with large good hysteresis.

We shall group Jelly Bean, Circle and Nearly BPS types together as type A or curved behaviour and group the Ohmic, SwitchOhmic and Nearly UPS types together as Type B memristors (i.e. those which are likely to give rise to type B or triangular behaviour over larger I-V range). Thus, we have 37.5\% are Type A and 57.8\% of the devices are Type B (29.7\% and 51.6\% respectively when the Nearly BPS and UPS are excluded as Type A and Type B respectively.) compared to 16\%: 83\% above, where we have included linear and triangular devices in type B (the difference is due to fabrication practicalities which meant that the sol mixture was only a day old before making the devices with different electrode sizes, but a week old before making those which were subjected to different gel forming preparations). The grouping into A and B type devices is based on the device's observed behaviour and the current range, see figure~\ref{fig:Types} and table~\ref{tab:DiffElecTypes} which shows that the different behaviours have different current ranges. The fact that type B devices have a HRS which is ohmic and highly conductive but an LRS which is of the same range as type A strongly suggests that type B devices work by a filament breaking and reforming mechanism and type A do not.

\begin{table}[htpn]
 \begin{tabular}{|c|c|l|l|l|l|l|}
  \hline
	Type	& Type of		& 1mm	& 2mm	& 3mm	& 4mm	& 5mm	\\
	No.	& Behaviour		&	&	&	&	&	\\
  \hline
	1 	& Ohmic 		&  4	& 5	& 8	& 1	& 6	\\			
	2	& Switch Ohmic		&  1	& 3	& 3	& -	& 3	\\
	3	& Jelly Bean		&  3	& 3	& 2	& 5	& 3	\\
	4	& Circle		&  1	& 1	& 1	& -	& -	\\
	5	& Nearly UPS		&  -	& -	& 2	& 1	& - 	\\ 
	6	& Nearly BPS		&  1	& 1	& -	& 1	& 1	\\
  \hline
	1+2	& Group B		& 50\% 	& 61\%	& 68\%	& 12\%	& 69\%	\\
	3+4	& Group A		& 40\%	& 30\%	& 19\%	& 62\%	& 23\%	\\
  \hline
	$\sum$ 	& All Types		& 10	& 13	& 16	& 8	& 13	\\
	\hline
 \end{tabular}
  \label{tab:SizeEffectDiffTypes}
  \caption{Testing the effect of electrode size on device behaviour selection. The data shows that there is no real effect of electrode size on behaviour selection. Note that the 4mm set was much smaller than the other datasets and the other batches of 4mm devices showed a preference for Ohmic-like behaviour over BPS-like.}
\end{table}

We looked in vain for a correlation between electrode size and a favouring of one type of behaviour over another, see table~\ref{tab:SizeEffectDiffTypes}. Essentially we generally get more of type B than type A (and here we've dropped the Nearly UPS and Nearly BPS from the analysis as these devices were not used for any further experiments). There is generally more type B than type A except for 4mm which was the smallest set tested (see the last line of table~\ref{tab:SizeEffectDiffTypes}), however the R and V series which were all 4mm wide had 66.7\% triangular to 16\% curved, suggesting that this result is due to the non-statistical sample sizes. There is a slight increase in the type B memristors with an increase in electrode size, which makes sense as a larger active areas allows a higher probability that a filament will have formed between the electrodes, however because of the results for the 4mm case, we suggest that there is not enough evidence to conclusively suggest that there is a favouring of type B memristors over other types as the active area size is increased. 

\begin{figure}[htbp!]
 \centering
 \includegraphics[width=0.75\textwidth]{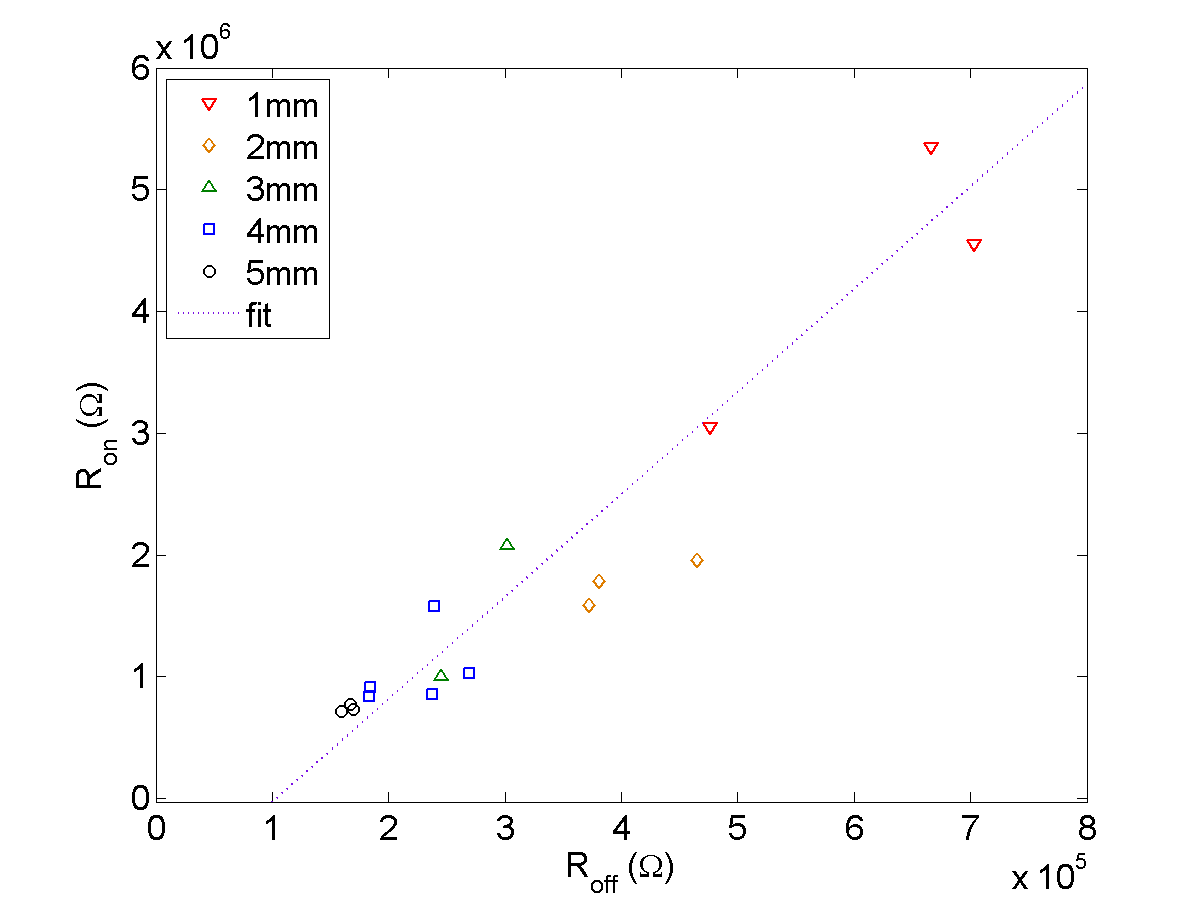}
 \caption{The effect of electrode width on the maximum and minimum resistances for the BPS-like memristors. There is a clear correlation where the larger electrodes leads to a lower resistance device.}
 \label{fig:RonVsRoff}
\end{figure}

\begin{figure}[htbp!]
 \centering
 \includegraphics[width=0.75\textwidth]{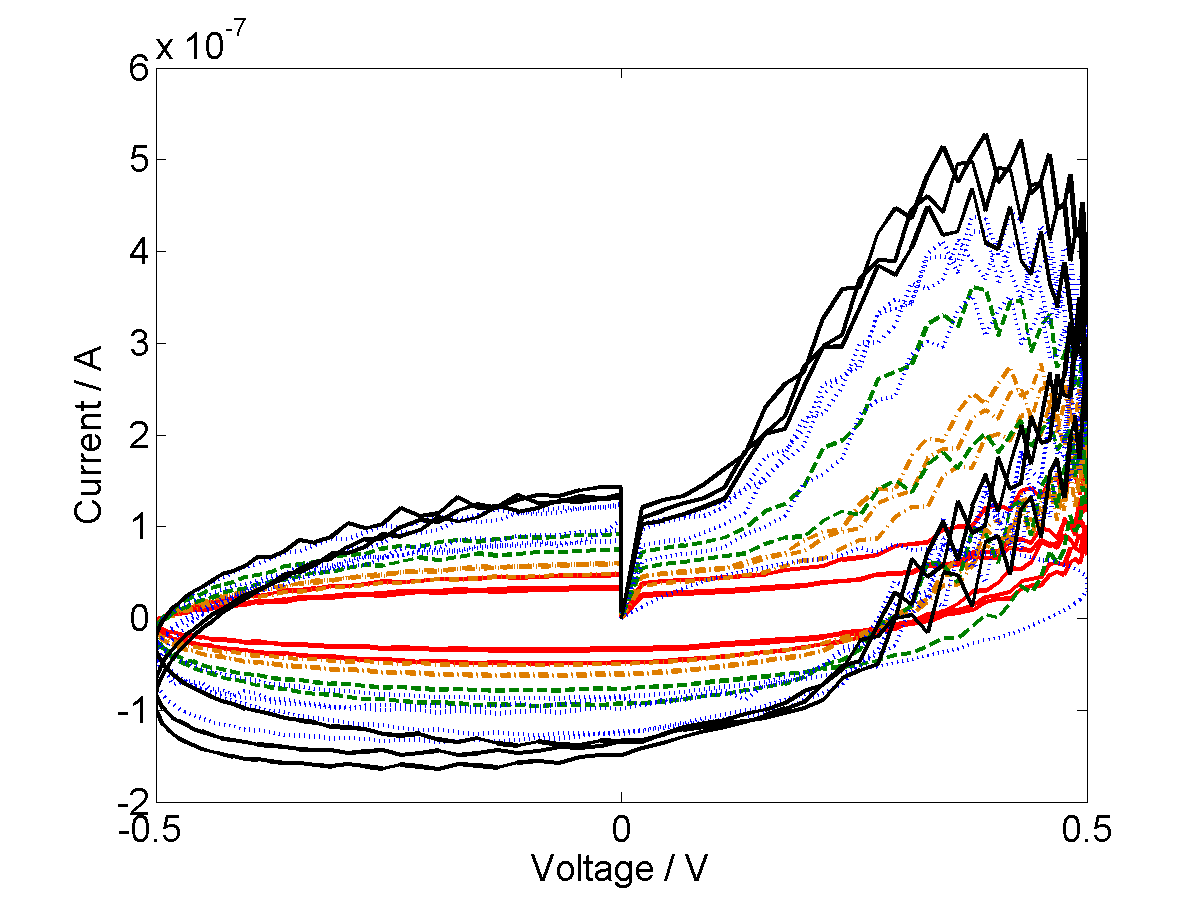}
 \caption{The effect of electrode size on memristor I-V curves. The larger the electrode, the bigger the I-V curve. Red = 1mm, Orange dot-dashes =2mm, blue dots = 3mm, green dashes = 4mm and black = 5mm.}
 \label{fig:JellyBean}
\end{figure}

There was no correlation between where the `Jelly Bean' curves cross zero and electrode width. There is a correlation between the resistance of Jelly Bean devices and the electrode width, as shown in figure~\ref{fig:RonVsRoff} and figure~\ref{fig:JellyBean}. As the electrode area is increased, it makes sense that current passed should be higher if the conductance is due to a bulk mechanism. If the conduction was due to connected filaments then there would be no relation between the resistance of the device and the electrode size, and this is exactly what we see for the Ohmic or SwitchOhmic devices. 

\subsection{Al-TiO$_2$-Au Memristors: The Effect of Aluminium}

12 Al-TiO$_2$-Au devices were run, 6 with the gold (top) electrode earthed, 6 with the aluminium (bottom) electrode earthed. As a comparison, the devices were run a second time with the wires switched (i.e. the aluminium electrode was earth). The virgin run did change the devices drastically, so the comparisons between virgin runs and repeated run are separated. 

\paragraph{Aluminium Electrodes are Necessary for Appreciable Memristance}

As would be expected if aluminium was essential to the operation of the flexible memristors, only one quadrant of the I-V curves showed appreciable memristance. Both types of behaviour, curved and triangular were seen. The current was in the range $10^{-4}A$ for curved and $10^{-3}A$ for triangular which is similar to Al-TiO$_2$-Al memristors. On their virgin run $\frac{10}{12}$ memristors are very asymmetrical (as defined as no hysteresis visible on the `off' side when the results are plotted over the full current range), $\frac{8}{12}$ on the second run. The `non-asymmetrical' devices have some hysteresis on both sides but usually there is still visible asymmetry. The non-hysteretic part of the very asymmetrical devices was investigated to find if the hysteresis was absent or merely much smaller. When the Au electrode was earthed the currents were only on the order of 10$^{-7}$-10$^{-8}$A with little discernible hysteresis, when the Al electrode was earthed two things were observed, a self-crossing hysteresis in the same current as above, and, in some devices, occasional current spikes of 10$^{-6}$A.

For the asymmetric virgin runs, 5/7 devices have memristance in the positive quadrant of I-V space when the Al electrode is earthed (Al is negative with respect to Au) and 6/6 have memristance in the negative quadrant when the Au electrode is earthed (Al is negative with respect to Au). Taking both virgin and second runs together, these numbers are 9/12 devices and 10/12. There is no memristance in the virgin runs when the Al electrode is positive with respect to the Au electrode. So, for appreciable memristance, the aluminium electrode has to be at a negative potential compared to the gold electrode.

\subsubsection{The Electrochemistry of Au-TiO$_2$-Al Memristors and What This Tells Us About Al-TiO$_2$-Al Memristors}

\begin{figure}[htbp]
 \centering
 \includegraphics[bb=0 0 680 462,scale=0.5,keepaspectratio=true]{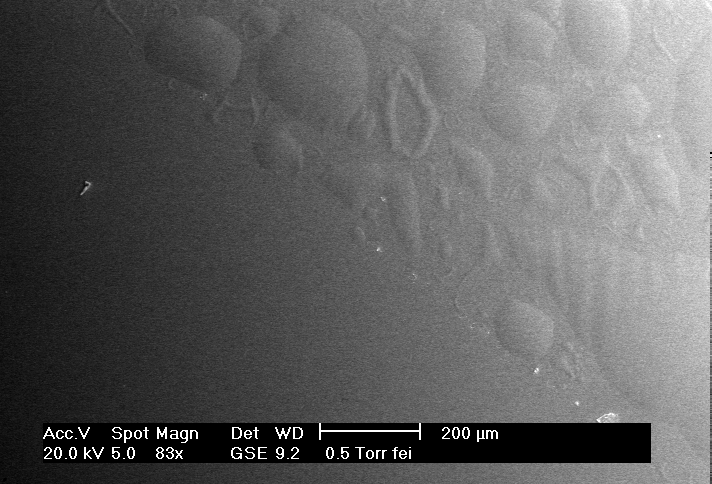}
 \caption{An SEM micrograph of the edge of the active area showing the sharp dividing line between the deformed electrode over the active area and the rest of the electrode.}
 \label{fig:EdgeOfBubbles}
\end{figure}

\begin{figure}[htbp!]
 \centering
 \includegraphics[scale=0.25,keepaspectratio=true]{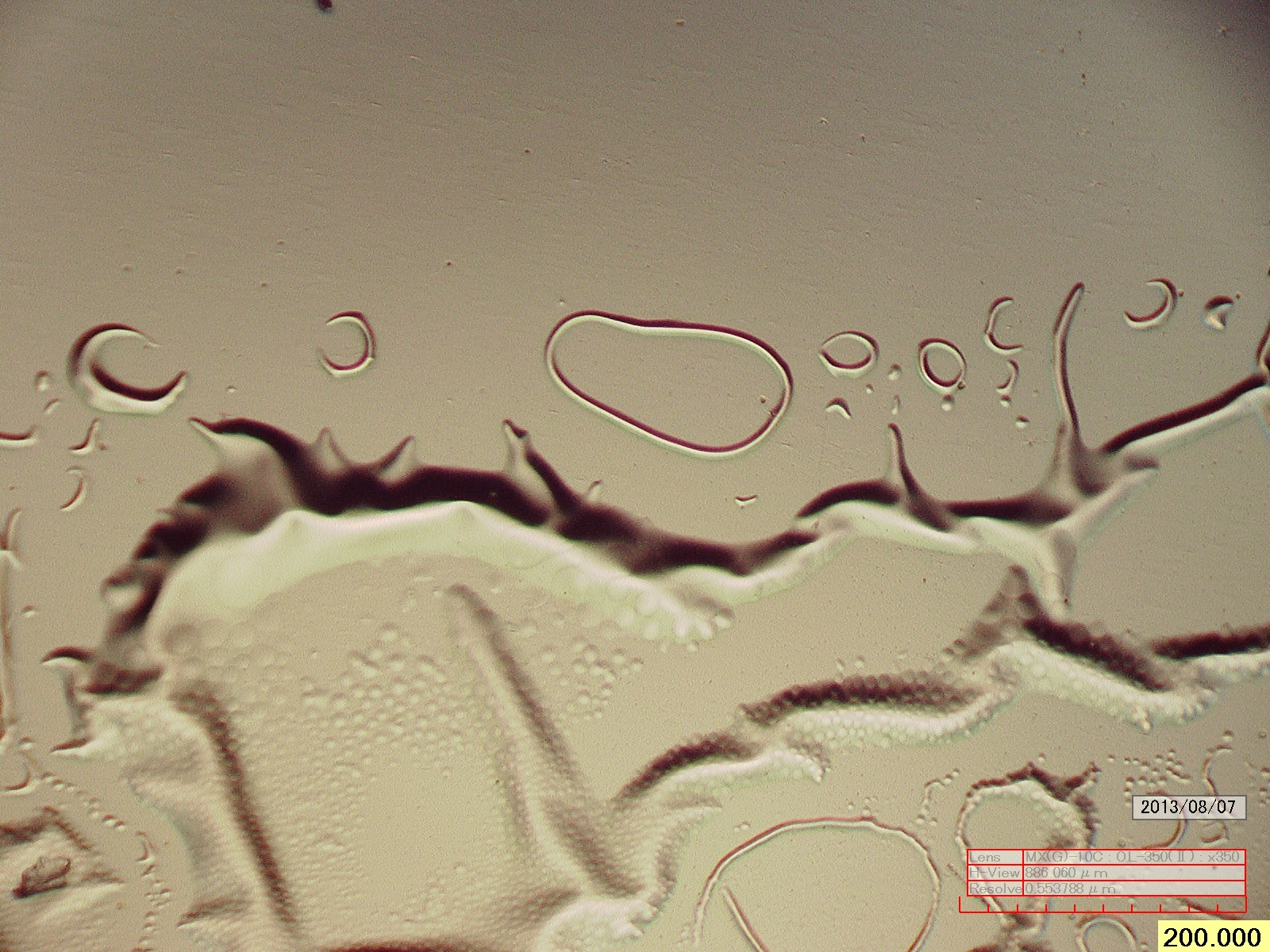}
 \caption{A side-lit light micrograph showing the different sizes of O$_2$ bubbles in the Al-TiO$_2$-Au devices. [Au14]}
 \label{fig:AuSideLitEdge}
\end{figure}

However, the Al-TiO$_2$-Au memristor cannot simply be viewed as half a memristor, it is a different system and has different electrochemistry. It was observed that the Al-TiO$_2$-Au devices changed in appearance after the first run (on a virgin device). It appeared as if the gold top electrode were melting and bubbling. These bubbles are shown in the SEM micrographs in figures~\ref{fig:FeildOfSurface} and ~\ref{fig:EdgeOfBubbles}. Some of the bubbles appear to have popped, see~\ref{fig:EdgeOfBubbles}. The photomicrograph in figure~\ref{fig:AuSideLitEdge} shows the sharpness of this edge.

\begin{figure}[htbp]
 \begin{tabular}{|c|c|c|c|c|}
  \hline
 \centering
 \includegraphics[bb=0 0 78 76]{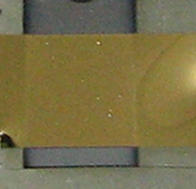} &
  \includegraphics[bb=0 0 78 76]{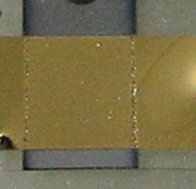} &
\includegraphics[bb=0 0 78 76]{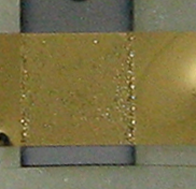} &
\includegraphics[bb=0 0 78 76]{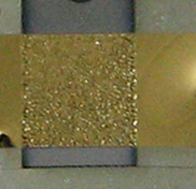} &
\includegraphics[bb=0 0 78 76]{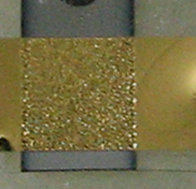} \\
  \hline
 \end{tabular}
  \label{fig:Au12}
  \caption{Stills from the movie of the Au12 undergoing oxygen evolution.}
\end{figure}

\begin{figure}[htbp]
 \begin{tabular}{|c|c|c|c|}
  \hline
 \centering
 \includegraphics[bb=0 0 78 76]{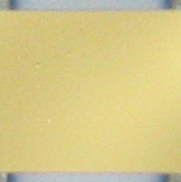} &
  \includegraphics[bb=0 0 78 76]{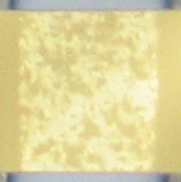} &
\includegraphics[bb=0 0 78 76]{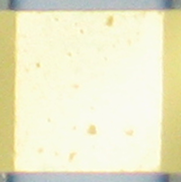} &
\includegraphics[bb=0 0 78 76]{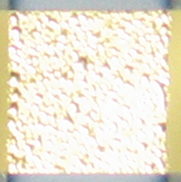}\\
  \hline
-1.3V	& -1.4V		& -1.5V		&-1.6V	\\
\hline
 \end{tabular}
  \label{fig:Au14}
  \caption{Stills from the movie of the Au12 undergoing oxygen evolution.}
\end{figure}

Two devices, AlAu-12 and AlAu-14, were connected taken through virgin run I-V cycles and photographed at every time step ($\pm0.1V$). As shown in figures~\ref{fig:Au12} and~\ref{fig:Au14}. When the gold electrode of AlAu-12 was earthed the transition from a smooth surface to a bubbled one happens at +1.5V ($\pm0.5V$), when the Au electrode is -1.5V with respect to the Aluminium electrode. For AlAu-14, when the aluminium electrode is earthed, the transition happens at -1.5V, again the Au electrode is negative with respect to the aluminium electrode. Note, this is the opposite way round to the appreciable memristance. In other words, if the Al is negative compared to Au we will usually see appreciable hysteresis, if the Au is negative with respect to the Al (the other half of the I-V curve), we will see very little hysteresis and instead the gold surface will deform. There is no feature in the I-V curve associated with this surface deformation at $\approx +1.5V$ (thus far we've always taken the source as being connected to the bottom electrode, following this schema the deformation happens in the positive quadrant of the I-V curve).

Both sets of photographs were lit slightly differently, AlAu-12 was side-lit with glancing light which serves to show up the 3-D structure of the surface, see figure~\ref{fig:Au12}, AlAu-14 was directly lit which makes the photos brighter and was strong enough light to allow us to see through the gold electrode film~\ref{fig:Au14}. AlAu-14 shows a change in structure at -1.4V before the surface bubbles are observed. This disturbance starts in the centre of the active area, but in both AlAu-12 and AlAu-14 the deformation of the gold electrode starts at the edge of the active area, where the top electrode would be under the largest mechanical stress. The light source was not moved during the photography and thus a very simple analysis of the photographs was possible. The photographs were in aligned in Photoshop and cropped to region shown in figure~\ref{fig:Au14}. These photographs were read into MatLab and the column brightness was averaged in each frame (similar results were seen when the row brightness was averaged). The blue channel was chosen to analyse as blue is the complimentary colour of yellow, the main hue of gold and thus the effect was more visible in this channel (NB. the file was in RGB format). Qualitatively similar but quantitatively smaller effects were seen in the red and green channels (yellow is a mix of red and green in this colourspace). The effect of this simple analysis matches what is seen with the eye, but adds two facts: the -1.4V change, coloured in blue happens between pixels 20 and 180 (this is inside the active area) and then the change in surface texture happens quickly and then stays relatively constant. Due to this irreversible change in electrode structure, Al-TiO$_2$-Au devices are not considered to be reproducible or particularly useful. Given the high availability costs of fabrication and uselessness as devices, the reverse (gold bottom electrode and aluminium top electrode) devices were not 
investigated.

This observed physical effect (bubble) can give information about the mechanism and structure of the memristors. Oxygen evolution has been observed in ReRAM, but not in gel-form TiO$_2$ memristors. Both the memristor and ReRAM communities agree that the resistance switching/memristance happens when the titanium dioxide is reduced from the stoichiometric TiO$_2$ form to an oxygen-deficient form TiO$_{(2-x)}$. TiO$_2$ is a semi-conductor. TiO$_{(2-x)}$ can be thought of as either have more free electrons (the electrochemist's viewpoint) or as having positive holes (the semiconductor physicist's viewpoint) or oxygen vacancies (the crystallographic chemist's viewpoint). Regardless of terminology, the reduced form of titanium dioxide is more conductive . There is debate about the exact form of the reduced form titanium dioxide (i.e. what the number $x$ is), it was thought to be a Rutile-based phase, but Magn\'{e}li phases have recently been implicated (see Introduction). 

The electrochemical mechanism, as taken from~\cite{155}, is as follows. The oxygen anions move under the effect of voltage (this is equivalent to the drift of oxygen vacancies mentioned in~\cite{F1}). Ti cations trap electrons emitted from the cathode thus:

$$
ne^{-} + \mathrm{Ti}^{4+} \longrightarrow \mathrm{Ti}^{(4-n)+}
$$

and there is oxidation at the anode (as this is where the oxygen anions are moving towards), possibly something like this, 

$$
\frac{2}{3} \mathrm{Al} + \mathrm{O} \longrightarrow \ddot{V}_{\mathrm{O}} + 2 e^{-} + \frac{1}{3} \mathrm{Al}_2\mathrm{O}_{3} \:.
$$

If the anode is gold or another noble metal, i.e. one that cannot form an oxide, we would get 

$$
\mathrm{O} \longrightarrow \ddot{V}_{\mathrm{O}} + 2 e^{-} + \frac{1}{2} \mathrm{O}_{2} \uparrow
$$

at the anode, where $\ddot{V}_{\mathrm{O}}$ is understood to be an oxygen vacancy in the TiO$_2$(of charge 2). ($\mathrm{O}_{2} \uparrow$ is the evolution of oxygen gas). As the gold is thin and malleable enough to deform, oxygen gas evolution would explain the gold electrode deformation. 

When we reverse the voltage polarity, i.e. when aluminium is the anode, we see no obvious gas evolution, no visible electrode deformation and memristive hysteresis in the I-V curve. This strongly suggests that the aluminium electrode is acting as a sink of oxygen anions. It is possible that in the Al-TiO$_2$-Al devices when the voltage direction is switched, the aluminium oxide layer is capable of acting as a source of oxygen anions. It is suggestive that the Al-TiO$_2$-Al devices are reproducible, in that they can be taken around the I-V curve several times, and the Al-TiO$_2$-Au devices are destroyed after the first run.

\begin{figure}[!tbp]
\centering
\subfigure[]{\includegraphics[width=0.49\textwidth]{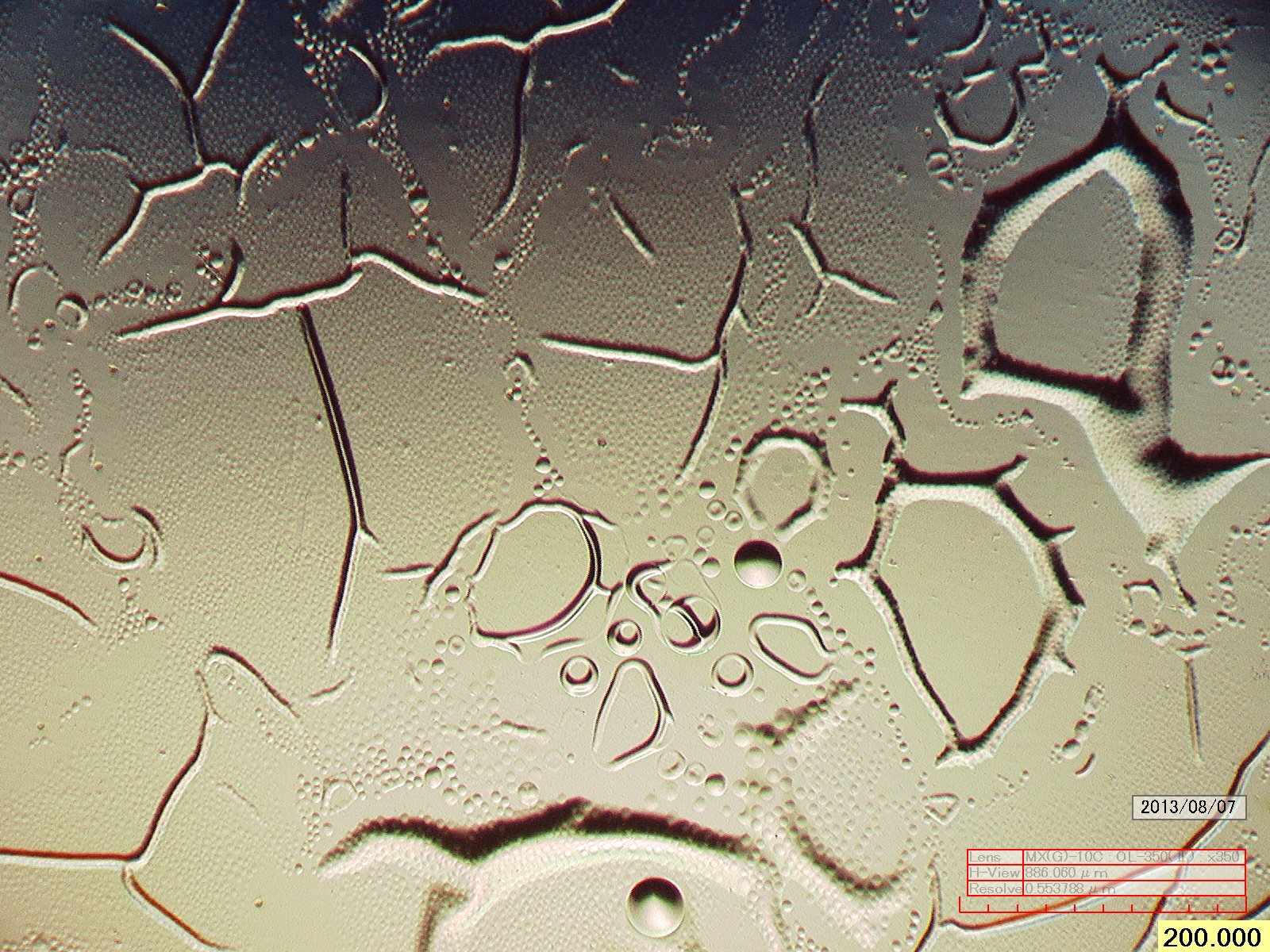}}
\subfigure[]{\includegraphics[width=0.49\textwidth]{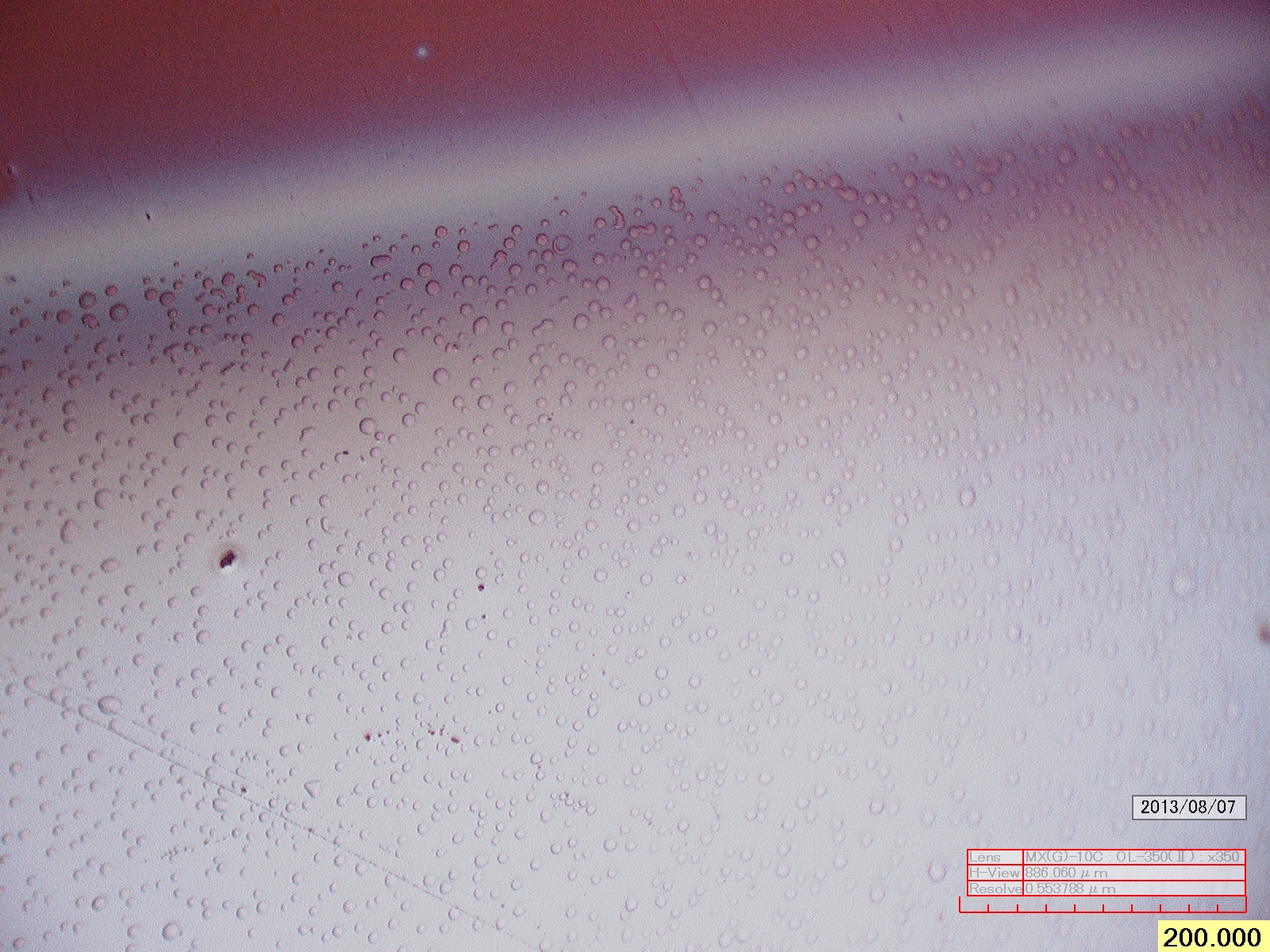}}
\caption{Different classification of device behaviour: a. A side-lit light micrograph showing the different sizes of O$_2$ bubbles in the Al-TiO$_2$-Au devices. [Au14]; b. An Al-TiO$_2$-Al micrograph after forming showing the presence of distortion in the top level of the electrodes caused by gas escape. This figure shows the edge of the active area.}
\label{fig:Bubbles}
\end{figure}

After observation of this electrode deformation effect for Au-TiO$_2$-Al memristors we put the Al-TiO$_2$ electrodes under the microscope. It had been observed by eye that the Al-TiO$_2$-Al memristors do look slightly cloudier after their virgin runs. Figure~\ref{fig:Bubbles} shows a comparison for the gold surface electrode (actually device AlAu-14) compared to an aluminium top electrode from a Al-TiO$_2$-Al device (V series). Although the top electrode is deformed after it's virgin run, the size of the bubbles is much smaller, specifically no larger around 11$\mu$m, whereas the bubble in figure~\ref{fig:AuSideLitEdge} is 556.25$\mu$m on its largest diameter. 


\begin{figure}[htbp]
 \centering
 \includegraphics[bb=0 0 680 462,scale=0.5,keepaspectratio=true]{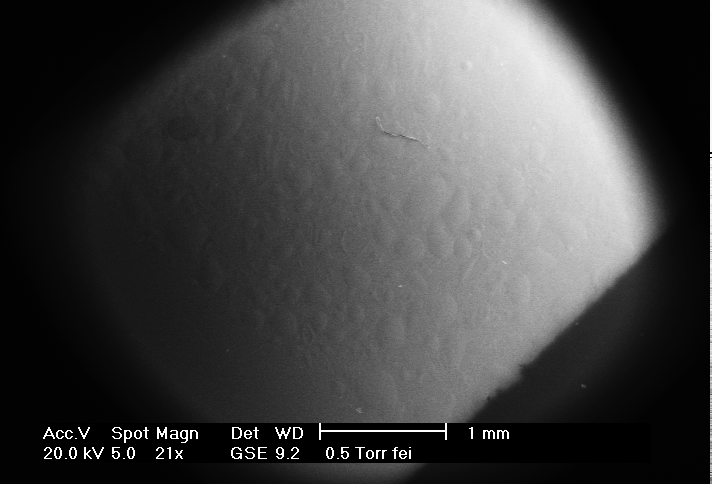}
 \caption{An SEM micrograph that shows that the bubbles are visible at the 1mm range and therefore resolvable by the human eye. We can see here that they bubbles cover the entire surface of the top electrode over the active area.}
 \label{fig:WideBubble}
\end{figure}

\clearpage

\clearpage

\section{Conclusion}

\begin{table}[htbp]
\begin{tabular}{|c|c|c|c|c|c|}
\hline
		& \multicolumn{3}{c|}{Aluminium} 		& \multicolumn{2}{c|}{Gold}	\\
\hline
		& 			& V series& R series	& 			& Al-TiO$_2$-Au 	\\
\hline
Aluminium	& UPS-like 		& 5	& 3		& UPS-like		& 6	\\
		& BPS			& 2	& 2		& BPS			& 2	\\
		& Mixed (BPS\&UPS)	& 2	& 4		& Mixed (BPS\&UPS)	& 1	\\
		& $-$			& $-$	& $-$		& Mixed (HT\&BPS)	& 2	\\
		& Half-triangle		& 2	& 3		& Half-triangle		& 1	\\
		& Total			& 12	& 12		& Total			& 12	\\
\hline
		& \multicolumn{3}{|c|}{ } 			& 			& Au-TiO$_2$-Au\\
\hline
		& \multicolumn{3}{|c|}{ }  			& BPS			& $-$	\\
Gold		& \multicolumn{3}{|c|}{ } 			& UPS-like		& $-$	\\
		& \multicolumn{3}{|c|}{ }  			& BPS			& $-$	\\
		& \multicolumn{3}{|c|}{ } 			& Mixed (BPS\&UPS)	& $-$	\\
		& \multicolumn{3}{|c|}{ }  			& Mixed (HT\&BPS)	& $-$	\\
		& \multicolumn{3}{|c|}{ }  			& Half-triangle		& 6	\\
		& \multicolumn{3}{|c|}{ }  			& Total			& 6	\\
\hline
\end{tabular}
\label{tab:SummaryTable}
\caption{Types of I-V curves seen with different electrodes. V series and R series refers to whether the bottom electrode was exposed to vacuum or air after sputtering. For the Al-TiO$_2$-Au the aluminium electrode was kept under vacuum. These were run with effectively no compliance current (it was set to 1A, a very high and non-limiting value). These figures are for virgin runs only. }
\end{table}

In order to elucidate how to control memristor electronic properties we have undertaken a study of several variable fabrication parameters, namely electrode material, humidity during gel formation and electrode size, and one variable measurement/forming parameter, namely the compliance current. 

We have found that using gold electrodes instead of aluminium does not give good memristor devices, instead we get ohmic low resistance resistors consistent with gold metal filaments, which may fuse to lower resistance resistors usable for WORMS memory or may switch to low-quality curved type A memristor behaviour consistent with the TiO$_2$ gel layer switching.  

Mixed electrode devices with an aluminium bottom electrode and a gold top electrode seem to suffer from oxygen evolution when gold is the anode and memristance when aluminium is the anode, suggesting that the aluminium or aluminium oxide layer on the electrode is involved in supporting memristance in flexible TiO$_2$ devices. These devices are useless as memristors or memory. 

Aluminium electrode devices (Al-TiO$_2$-Al) offer a richer behaviour which we have classified into two types. Type A or curved behaviour is explainable using fundamental memristor theory, specifically they are non-transversal memristors (see~\cite{276}) when measured at low voltages and appear to be transversal memristors when measured at larger voltages. Type A behaviour is also describable as ReRAM BPS. Changing humidity during sol formation, compliance current or electrode size does not increase the number of type A behaviour. The resistance of type A memristors can be reduced by increasing the electrode size which is consistent with type A behaviour being due to a bulk motion of oxygen vacancies. Although not presented here, we have found that type A memristors tend to be more reproducible on repeated runs and are useful for building neuromorphic circuits, logic gates and as memristors. 

Type B or triangular behaviour seems similar to ReRAM UPS behaviour, although as ReRAM researchers rarely subject their UPS devices to a bipolar voltage waveform as we have here we can't be sure. Technically type B behaviour does not fit the strict definition for the memristor~\cite{14,84} because the resistance of the LRS is not non-linear. This fact, taken with the facts that the resistance of the LRS state is not related to the size of the electrode strongly suggests that the LRS is due to a filament which forms and is then broken and reformed (a fuse, anti-fuse mechanism). We can't definitively conclude here that the filament is due to oxygen vacancy motion and not a titanium cation wire, however the existence of gas evolution bubbles suggests that this is the case. As the LRS state is curved it suggests that type A behaviour is there alongside type B behaviour, in that the effects of the bulk movement of oxygen vacancies is drowned out by the large ohmic current until the filament breaks. This is supported by the mixed devices which switch the type A behaviour after the filament breaks. Extending the basic memristor theory to include a filament would allow the device to be described by memristor-based equations. Taking a wider definition of memristance such as that described in~\cite{119} includes these devices as memristive systems. To get the largest number type B devices we want to allow our gel to form under vacuum and use a compliance current to prevent irreversible anti-fusing. Although not presented here, we have found that type B devices are less reproducible on repeated runs (as the filament does not break at exactly the same voltage it breaks at during the virgin run, which is consistent with results for ReRAM UPS), and seem to be useful as memory or as neuromorphic circuits.

\section*{References}

\bibliography{UWELit}



\end{document}